\documentclass[prd,aps,showpacs,nofootinbib,preprint,tightenlines]{revtex4}

\newcommand{\checked}[1]{}

\usepackage{graphicx}
\usepackage{rotating}
\usepackage{axodraw}

\newcommand{\beq}{\begin{equation}}
\newcommand{\eeq}{\end{equation}}
\newcommand{\bqa}{\begin{eqnarray}}
\newcommand{\eqa}{\end{eqnarray}}

\newcommand{\picb}[1]{\;\parbox[c]{48pt}{\begin{picture}(45,30)(-9,0)
\SetWidth{1.0}\SetScale{1.0} #1 \end{picture}}\;}

\def\Lwidth{1}

\def\Agl(#1,#2)(#3,#4,#5){\PhotonArc(#1,#2)(#3,#4,#5){\Lwidth}
{6.283 #3 mul 360 div #4 #5 sub #4 #5 sub mul sqrt mul Ldensity mul}}
\def\Lgl(#1,#2)(#3,#4){\Photon(#1,#2)(#3,#4){\Lwidth}
{#1 #3 sub #1 #3 sub mul #2 #4 sub #2 #4 sub mul add sqrt Ldensity mul}}
\def\Agh(#1,#2)(#3,#4,#5){\DashArrowArc(#1,#2)(#3,#4,#5){1}}
\def\Aagh(#1,#2)(#3,#4,#5){\DashArrowArcn(#1,#2)(#3,#5,#4){1}}
\def\Lgh(#1,#2)(#3,#4){\DashArrowLine(#1,#2)(#3,#4){1}}
\def\Lagh(#1,#2)(#3,#4){\DashArrowLine(#3,#4)(#1,#2){1}}
\def\Ahh(#1,#2)(#3,#4,#5){\DashCArc(#1,#2)(#3,#4,#5){1}}
\def\Lhh(#1,#2)(#3,#4){\DashLine(#1,#2)(#3,#4){1}}
\def\Aqu(#1,#2)(#3,#4,#5){\ArrowArc(#1,#2)(#3,#4,#5)}
\def\Aaqu(#1,#2)(#3,#4,#5){\ArrowArcn(#1,#2)(#3,#5,#4)}
\def\Lqu(#1,#2)(#3,#4){\ArrowLine(#1,#2)(#3,#4)}
\def\Laqu(#1,#2)(#3,#4){\ArrowLine(#3,#4)(#1,#2)}
\def\Aqq(#1,#2)(#3,#4,#5){\CArc(#1,#2)(#3,#4,#5)}
\def\Lqq(#1,#2)(#3,#4){\ArrowLine(#1,#2)(#3,#4)}
\def\Asc(#1,#2)(#3,#4,#5){\ArrowArc(#1,#2)(#3,#4,#5)}
\def\Lsc(#1,#2)(#3,#4){\ArrowLine(#1,#2)(#3,#4)}
\def\DAsc(#1,#2)(#3,#4,#5){\DashCArc(#1,#2)(#3,#4,#5){3}}
\def\DLsc(#1,#2)(#3,#4){\DashLine(#1,#2)(#3,#4){3}}
\def\TAsc(#1,#2)(#3,#4,#5){\SetWidth{2.0}\CArc(#1,#2)(#3,#4,#5)\SetWidth{1.0}}
\def\TLsc(#1,#2)(#3,#4){\SetWidth{2.0}\ArrowLine(#1,#2)(#3,#4)\SetWidth{1.0}}

\begin{document}

\title{Energy Loss of a Heavy Fermion in an Anisotropic QED Plasma}

\preprint{ TUW-03-25 }

\author{Paul Romatschke}
\author{Michael Strickland}
\affiliation{Institut f\"ur Theoretische Physik, Technische Universit\"at Wien,
	Wiedner Hauptstrasse 8-10, A-1040 Vienna, Austria
     \vspace{1cm}
	}

\begin{abstract}
We compute the leading-order collisional energy loss of a heavy fermion 
propagating in a QED plasma with an electron distribution function which is 
anisotropic in momentum space. We show that in the presence of such anisotropies 
there can be a significant directional dependence of the heavy fermion energy 
loss with the effect being large for highly-relativistic velocities.  We 
also repeat the analysis of the isotropic case more carefully and show that the 
final result depends on the intermediate scale used to separate hard and soft 
contributions to the energy loss.  We then show that the canonical isotropic result is
obtained in the weak-coupling limit.  For intermediate-coupling we use the
residual scale dependence as a measure of our theoretical uncertainty.  
We also discuss complications which could arise due to the presence of 
unstable soft photonic modes and demonstrate that the calculation of the energy
loss is safe. 
\end{abstract}
\maketitle
\newpage

\small

\section{Introduction}

An understanding of the production, propagation, and hadronization of heavy 
quarks in relativistic heavy ion collisions is important for predicting a number 
of experimental observables including the heavy-meson spectrum, the single 
lepton spectrum, and the dilepton spectrum.  The first experimental results for 
the inclusive electron spectrum have been reported \cite{adcox:2003} in addition 
to the first measurements of $J/\psi$ production at RHIC \cite{nagle:2002}.  The 
measurement of the inclusive electron spectrum allows for a determination of 
heavy quark energy loss since it is primarily due to the semi-leptonic decay of 
charm quarks.  The heavy fermion energy loss comes into play since it is 
necessary in order to predict the heavy fermion energy at the decay point. It is 
therefore important to have a thorough theoretical understanding of heavy 
fermion energy loss for a proper comparison with the experimental results.  In 
this paper we will show that in QED there is a modification of the leading-order 
(collisional) heavy fermion energy loss if there is a momentum-space anisotropy 
in the electron distribution function. The motivation for this work is to 
provide a testing ground for the techniques which can be applied to QCD in order 
to make predictions of the directional dependence of the collisional energy loss 
of a heavy fermion propagating through an anisotropic quark-gluon plasma. 

In the last few years a more or less standard picture of the early stages of a 
relativistic heavy ion collision has emerged.  In its most simplified form 
there are three assumptions: (1) that the system is boost invariant along the 
beam direction, (2) that it is homogeneous in the directions perpendicular to the 
beam direction, and (3) that the physics at early times is dominated by gluons 
with momentum at a ``saturation'' scale $Q_s$ which have occupation numbers of 
order $1/\alpha_s$. The first two assumptions are reasonable for describing the 
central rapidity region in relativistic heavy-ion collisions.  The third 
assumption relies on the presence of gluonic ``saturation'' of the nuclear 
wavefunction at very small values of the Bjorken variable $x$ \cite{saturation}. 
In this regime one can determine the growth of the gluon distribution by 
requiring that the cross section for deep inelastic scattering at fixed $Q^2$ 
does not violate unitarity bounds.  As a result the gluon distribution 
function saturates at a scale $Q_s$ changing from $1/k_\perp^2 \rightarrow 
\log(Q_s^2/k_\perp^2)/\alpha_s$.  Luckily, despite this saturation, due to the 
factor of $1/\alpha_s$ in the second scaling relation the occupation number of 
small-$x$ gluonic modes in the nuclear wavefunction is still large enough to 
determine their distribution function analytically using classical nonlinear 
field theory \cite{saturation}.  In the weak-coupling limit the assumptions 
above have been used by Baier et al. in an attempt to systematically describe 
the early stages of quark-gluon plasma evolution in a framework called ``bottom-up'' 
thermalization \cite{BMSS:2001}.

The resulting picture which emerges from using these assumptions is one in which the 
initial gluonic distribution function is extremely anisotropic in momentum space 
having the form

\beq
 f({\bf p},{\bf x}) = F(p_\perp) \delta(p_z) \, .
\label{deltalimit}
\eeq
This is, of course, an idealization.  In a more realistic scenario the delta 
function above would have a small but finite width which increases as a function 
of time, but despite this finite width, the distribution function would still be 
extremely anisotropic in momentum space during the early stages of the 
collision.  In anisotropic systems it has been shown that the physics of the QED 
and QCD collective modes changes dramatically compared to the isotropic case and 
instabilities are present which can accelerate the thermalization and 
isotropization of the plasma 
\cite{Weibel:1959,SM:1993,SM:1994,SM:1997,RM:2003,RS:2003,ALM:2003}. In fact, 
the paper of Arnold, Lenaghan, and Moore points out that 
the presence of anisotropies 
in the early stages of QGP evolution requires modification of the ``bottom-up'' 
thermalization scenario \cite{ALM:2003}.  

In our previous paper \cite{RS:2003} we derived a tensor basis for the 
photon/gluon self-energy and calculated the corresponding structure functions 
for an anisotropic system in which the distribution function is homogeneous and 
obtained from an isotropic distribution function by the rescaling of only one 
direction in momentum space

\beq
f({\bf p},{\bf x}) = N(\xi)\,f_{\rm iso}\left(p\sqrt{1+\xi({\bf v} \cdot {\bf n})^2}\right) \, ,
\label{distfunc}
\eeq
where $N(\xi)$ is a normalization constant, ${\bf v}$ is the particle 3-velocity, 
${\bf \hat n}$ specifies the anisotropy direction, and $-
1<\xi<\infty$.  Note that the distribution function given by 
Eq.~(\ref{deltalimit}) is obtained in the limit $\xi\rightarrow\infty$ assuming 
that 
Eq.~(\ref{distfunc}) is normalized in the same way as Eq.~(\ref{deltalimit}).  

Using classical kinetic field theory we were then able to numerically determine 
the photon/gluon self-energy structure functions for this tensor basis in the 
entire complex energy plane for arbitrary $\xi$. In this paper we will use these 
structure functions to determine the leading-order (collisional) energy loss of 
a heavy fermion propagating through an anisotropic QED plasma.  This calculation 
will allow us to determine the dependence of the collisional energy loss on the 
angle of propagation $\theta_n$, the parton velocity $v$, the coupling constant 
$e$, and the temperature $T$. We will show that for large anisotropies and 
velocities the directional dependence of the heavy fermion energy loss is large 
and could therefore lead to a significant experimental effect. During the 
development we will also discuss some technical details related to the cutoff 
dependence of the isotropic and anisotropic results.

The organization of the paper is as follows.  In Sec.~\ref{setup} we review some 
of the notation and results from Ref.~\cite{RS:2003}.  In Sec.~\ref{softpart} we 
calculate the contribution to the energy loss coming from the exchange of soft 
photons with momenta on the order of $eT$.  In Sec.~\ref{sect:shielding} we 
discuss complications which could arise due to the presence of unstable soft
photonic modes and show that the soft energy loss calculation is safe.
In Sec.~\ref{hardpart} we calculate 
the contribution to the energy loss coming from the exchange of hard photons 
with momenta on the order of $T$.  In Sec.~\ref{results} we combine the soft and 
hard contributions to obtain the final isotropic and anisotropic results. In 
Sec.~\ref{conclusions} we list the limitations of the treatment presented here 
and provide a short description of how the result contained here can be extended 
to QCD.

\section{Setup}
\label{setup}

In this section we will review some of the findings from our previous paper 
\cite{RS:2003} which we will use in the subsequent sections.

\subsection{Tensor Basis}

For anisotropic systems with only one preferred direction we construct a basis 
for symmetric 3-tensors that depends on a fixed anisotropy 3-vector $n^{i}$ with 
$n^2=1$.  We first define the projection operator
\begin{equation}
A^{ij}=\delta^{ij}-k^{i}k^{j}/k^2,
\end{equation}
\checked{mp}
and use it to construct $\tilde{n}^{i}=A^{ij} n^{j}$ which obeys
$\tilde{n} \cdot k =0$. With this we can construct the tensors
\begin{equation}
B^{ij}=k^{i}k^{j}/k^2
\end{equation}
\begin{equation}
C^{ij}=\tilde{n}^{i} \tilde{n}^{j} / \tilde{n}^2
\end{equation}
\begin{equation}
D^{ij}=k^{i}\tilde{n}^{j}+k^{j}\tilde{n}^{i}.
\end{equation}
\checked{mp}
Any symmetric 3-tensor ${\bf T}$ can now be decomposed into the basis spanned by
the four tensors ${\bf A},{\bf B},{\bf C},$ and ${\bf D}$
\beq
{\bf T}=a\,{\bf A}+b\,{\bf B}+c\,{\bf C}+d\,{\bf D} \; .
\eeq
\checked{mp}

\subsection{Self-energy structure functions}

Converting the result of Ref.~\cite{RS:2003} from QCD to QED the spacelike 
components of the high-temperature photon self-energy for particles with soft 
momentum ($k \sim e T$) can be written as
\begin{equation}
\Pi^{i j}(K) = - 4 e^2 \int \frac{d^3{\bf p}}{(2\pi)^3} v^{i} \partial^{l} f_e({\bf p})
\left( \delta^{j l}+\frac{v^{j} k^{l}}{K\cdot V + i \epsilon}\right) \; ,
\label{selfenergy2}
\end{equation}
\checked{mp}
where the electron distribution function $f_e({\bf p})$ is, in principle, 
completely arbitrary.  In what follows we will assume that $f_e({\bf p})$ can be 
obtained from an isotropic distribution function by the rescaling of only one 
direction in momentum space.  In practice this means that, given any isotropic 
distribution function $f_{e,{\rm iso}}(p)$, we can construct an 
anisotropic version by changing the argument of the isotropic distribution 
function
\begin{equation}
f_e({\bf p}) = N(\xi) \, f_{e,{\rm iso}}\left(\sqrt{{\bf p}^2+\xi({\bf p}\cdot{\bf \hat n})^2}\right) \; ,
\label{distfunc2}
\end{equation}
\checked{mp}
where $N(\xi)$ is a normalization constant, ${\bf \hat n}$ is the direction of 
the anisotropy and $\xi>-1$ is a adjustable anisotropy parameter.  Note that 
$\xi>0$ corresponds to a contraction of the distribution along the ${\bf \hat n}$ 
direction whereas $-1<\xi<0$ corresponds to a stretching of the distribution along
the ${\bf \hat n}$ direction.

The normalization constant, $N(\xi)$, can be determined by normalizing the distribution
function to a fixed number density for all values of $\xi$
\bqa
\int_{\bf p} f_{e,{\rm iso}}(p) =  \int_{\bf p} f_e({\bf p}) 
	= N(\xi) \int_{\bf p} f_{e,{\rm iso}}\left(\sqrt{{\bf p}^2+\xi({\bf p}\cdot{\bf \hat n})^2}\right) \, \, .
\label{normcondition}
\eqa
By performing a change of variables to $\tilde p$
\begin{equation}
\tilde{p}^2=p^2\left(1+\xi ({\bf v}\cdot{\bf n})^2\right) \; ,
\label{varchange}
\end{equation}
\checked{mp}
on the right hand side we see that the normalization 
condition given in (\ref{normcondition}) requires that
\bqa
N(\xi) = \sqrt{1+\xi} \, .
\eqa

Making the same change of variables (\ref{varchange}) also in 
(\ref{selfenergy2}) it is possible to integrate out the $|\tilde p|$-dependence 
giving
\begin{equation}
\Pi^{i j}(K) = m_{D}^2 \sqrt{1+\xi} \int \frac{d \Omega}{4 \pi} v^{i}%
\frac{v^{l}+\xi({\bf v}.{\bf n}) n^{l}}{%
(1+\xi({\bf v}.{\bf n})^2)^2}
\left( \delta^{j l}+\frac{v^{j} k^{l}}{K\cdot V + i \epsilon}\right) \; ,
\end{equation}
\checked{mp}
where
\beq
m_D^2 = -{2 e^2\over \pi^2} \int_0^\infty d p \,
  p^2 {d f_{e,{\rm iso}}(p) \over dp} \; .
\eeq
\checked{m}
Note that in the analysis which follows we will assume that $f_{e,{\rm iso}}(p) = 
n_F(p)=(\exp({p/T})+1)^{-1}$, in which case $m_D^2 = e^2 T^2/3$. 

We can then decompose the self-energy into four structure functions
\begin{equation}
{\bf \Pi}= \alpha\,{\bf A} + \beta\,{\bf B} + \gamma\,{\bf C} + \delta\,{\bf D} \; ,
\end{equation}
\checked{mp}
which are determined by taking the following contractions:
\begin{eqnarray}
\hat{\bf k}\cdot{\bf\Pi}\cdot\hat{\bf k} & = & \beta \; , \nonumber \\
\tilde{\bf n}\cdot{\bf\Pi}\cdot\hat{\bf k} & = & \tilde{n}^2 k \delta \; , \nonumber \\
\tilde{\bf n}\cdot{\bf\Pi}\cdot\tilde{\bf n} & = & \tilde{n}^2 (\alpha+\gamma) \; , \nonumber \\
{\rm Tr}\,{\bf \Pi} & = & 2\alpha +\beta +\gamma \; .
\label{contractions}
\end{eqnarray}
\checked{mp}

All four structure functions depend on $m_D$, $\omega$, $k$, $\xi$, and
${\bf \hat k}\cdot{\bf n}$.
In the limit $\xi \rightarrow 0$ the structure functions $\alpha$ and $\beta$
reduce to the isotropic hard-thermal-loop self-energies and $\gamma$ and $\delta$
vanish
\bqa
\lim_{\xi\rightarrow0}\alpha(K) &=& \Pi_T(K) + {\cal O}(\xi) \; , \nonumber \\
\lim_{\xi\rightarrow0}\beta(K) &=& {\omega^2\over k^2} \Pi_L(K) + {\cal O}(\xi) \; , \nonumber \\
\lim_{\xi\rightarrow0}\gamma(K) &=& {\cal O}(\xi) \; , \nonumber \\
\lim_{\xi\rightarrow0}\delta(K) &=& {\cal O}(\xi)\; ,
\label{isolimit}
\eqa
\checked{mp}
with
\bqa
\Pi_T(K) &=& {m_D^2\over2} {\omega^2 \over k^2}
  \left[1-{\omega^2-k^2 \over 2 \omega k}\log{\omega+k\over\omega-k}\right] \, , 
  \label{pit} \\
  \Pi_L(K) &=& m_D^2 \left[ {\omega\over 2k} \log{\omega+k\over\omega-k}-1\right] \; .
\label{pil}
\eqa
\checked{mp}
The ${\cal O}(\xi)$ terms in (\ref{isolimit}) were determined 
analytically in Ref.~\cite{RS:2003}.
\footnote{Note that the normalization used in Ref.~\cite{RS:2003} is different than the normalization
used here.  The weak-anisotropy expansions for the normalization contained here
can be obtained from the expressions contained in Ref.~\cite{RS:2003} by taking
$m_D^2\rightarrow\sqrt{1+\xi}\,m_D^2$ and re-expanding in $\xi$.}

With these structure functions in hand we can construct the propagator 
$\Delta^{ij}(K)$ using the expressions from the previous section
\beq
\Delta(K) =  \Delta_A \, [{\bf A}-{\bf C}] 
        + \Delta_G \, [(k^2 - \omega^2 + \alpha + \gamma) {\bf B} + 
        (\beta-\omega^2) {\bf C}  - \delta {\bf D}] \; ,
\eeq
\checked{mp}
where
\bqa
\Delta_A^{-1}(K) &=& k^2 - \omega^2 + \alpha \; , \label{propfnc1-2} \\
\Delta_G^{-1}(K) &=& (k^2 - \omega^2 + \alpha + \gamma)(\beta-\omega^2)-k^2 \tilde n^2 \delta^2 \; .
\label{propfnc2-2}
\eqa
\checked{mp}

\section{Energy loss calculation}

In this section we compute the collisional energy loss of a heavy fermion 
propagating through an electromagnetic plasma which has an anisotropic momentum 
space electron distribution function.  The starting point for the calculation is 
the same as in the isotropic case \cite{BT:1991}.  In these papers Braaten and 
Thoma calculated the energy loss of a heavy fermion for both QED and QCD.  Here 
we will concentrate on QED but the extension to QCD is relatively 
straightforward.  Note that there were previous calculations of the in-medium 
partonic energy loss \cite{JB:1982,GT:1990,SM:1991}; however, the calculation of 
Braaten and Thoma was the first to present a systematic method for performing 
the complete order $g$ calculation of this quantity.  

The technique used by Braaten and Thoma was to consider independently the 
contributions from soft photon exchange ($q \sim m_D \sim e T$) and hard photon 
exchange ($q \sim T$).  Each of these quantities is divergent with the soft 
contribution having a logarithmic UV divergence and the hard contribution having 
a logarithmic IR divergence.  In their paper Braaten and Thoma found that if an 
arbitrary momentum scale $q^*$ was introduced to separate the hard and soft 
regions then the dependence on this arbitrary scale vanished when the 
contributions are combined and the result obtained is finite.  

Below we will use the same technique as Braaten and Thoma but we will show that 
if the resulting integrals are treated more carefully then the final result 
obtained by adding the soft and hard contributions does depend on $q^*$. 
However, in the weak-coupling limit $e\ll 1$ the dependence on $q^*$ is small as 
long as $m_D \ll q^* \ll T$ and the Braaten-Thoma result is reproduced.  Since 
the full result does depend on $q^*$ we will need a prescription for fixing it. 
Here we will fix $q^*$ using the ``principle of minimal sensitivity'' which in 
practice means that we minimize the energy loss with respect to $q^*$ and 
evaluate at this point.  We can then obtain a measurement of our theoretical 
uncertainty by varying $q^*$ by a fixed amount around this point.  In the next 
three subsections we derive integral expressions for the soft and hard 
contributions to the energy loss and present some results for these in the limit 
of small anisotropies.  We also discuss complications which could arise due
to the presence of plasma instabilities. 

\subsection{Soft Part}
\label{softpart}

For calculating the soft energy loss one can use classical field theory methods. 
The classical expression for parton energy loss per unit of time is
\beq
\left({{\rm d} W\over {\rm d} t}\right)_{\rm soft} = {\rm Re} \, \int d^3{\bf x} \; {\bf J}_{\rm ext}(X)\cdot{\bf E}_{{\rm ind}}(X) \; ,
\label{eloss1}
\eeq
\checked{mp} 
where $X=(t,{\bf x})$, and ${\bf J}_{\rm ext}$ is the current induced by 
a test fermion propagating with velocity ${\bf v}$:
\bqa
{\bf J}_{\rm ext}(X) &=& e {\bf v} \delta^{(3)}({\bf x}-{\bf v}t) \; , \nonumber \\
{\bf J}_{\rm ext}(Q) &=& (2 \pi)  e {\bf v} \delta(\omega - {\bf q}\cdot{\bf v}) \; ,
\eqa
\checked{mp}
with $Q=(\omega,{\bf q})$.  Using
\beq
E^i_{{\rm ind}}(Q) = i \omega \, \left(\Delta^{ij}(Q)-\Delta_{0}^{ij}(Q)\right)%
 J_{\rm ext}^j(Q) \; ,
\eeq
where $\Delta^{ij}_0$ is the free propagator, 
and Fourier transforming to $X$ we obtain
\beq
E^i_{{\rm ind}}(X) = i e  \int {d^3 {\bf q} \over (2 \pi)^3} \, ({\bf q}\cdot{\bf v}) (\Delta^{ij}(Q)-\Delta^{ij}_0(Q)) v^j%
  e^{i({\bf q}\cdot{\bf x}-({\bf q}\cdot{\bf v}) t) } \; .
\eeq
\checked{mp}
The above equation allows us to write Eq.~(\ref{eloss1}) as
\beq
-\left({{\rm d} W\over {\rm d} t}\right)_{\rm soft} = e^2  \, {\rm Im} \, \int {d^3 {\bf q} \over (2 \pi)^3} \, ({\bf q}\cdot{\bf v}) v^i (\Delta^{ij}(Q)-\Delta^{ij}_0(Q)) v^j  \; .
\label{eloss2}
\eeq
\checked{mp}
The propagator can be expanded in our tensor basis and the contractions
on the right-hand-side give
\bqa
v^i A^{ij} v^j &=& v^2 - (\hat{\bf q}\cdot{\bf v})^2 \; , \nonumber \\
v^i B^{ij} v^j &=& (\hat{\bf q}\cdot{\bf v})^2 \, , \nonumber \\
v^i C^{ij} v^j &=& (\tilde{\bf n}\cdot{\bf v})^2/\tilde{n}^2  \, ,\nonumber \\
v^i D^{ij} v^j &=& 2 ({\bf q}\cdot{\bf v}) ( \tilde{\bf n}\cdot{\bf v})  \; ,
\eqa
so that together with $dW/dx=v^{-1} dW/dt$ we obtain
\bqa
-\left({{\rm d} W\over {\rm d} x}\right)_{\rm soft} &=& \frac{e^2}{v}  \, {\rm Im} \, \int {d^3 {\bf q} \over (2 \pi)^3} \, \omega \,
	 \left(\Delta_A(Q)-{1\over q^2-\omega^2}\right) \, \left[ v^2 - {\omega^2 \over q^2} - {(\tilde{\bf n}\cdot{\bf v})^2\over\tilde{n}^2} \right] \nonumber \\
       && \hspace{-7mm} + \,\omega \, \Delta_G(Q) \, \left[ {\omega^2\over q^2}(q^2 - \omega^2 + \alpha + \gamma) + (\beta-\omega^2) {(\tilde{\bf n}\cdot{\bf v})^2 \over \tilde{n}^2}  - 2 \delta \omega ( \tilde{\bf n}\cdot{\bf v}) \right]
	\; \nonumber \\
       && \hspace{-7mm} + \, {1\over\omega(q^2-\omega^2)} \, \left[ {\omega^2\over q^2}(q^2 - \omega^2) - \omega^2 {(\tilde{\bf n}\cdot{\bf v})^2 \over \tilde{n}^2}  \right]
	\; ,
\label{elossReg}
\eqa
\checked{p}
with $\omega = {\bf q}\cdot{\bf v}$.  Performing some algebraic transformations 
and scaling out the momentum gives
\bqa
\left({{\rm d} W\over {\rm d} x} \right)_{\rm soft} &=& \frac{e^2}{v}  
\, {\rm Im} \, \int {d^3 {\bf q} \over (2 \pi)^3} \, \frac{\hat{\omega}}{q (1-\hat{\omega}^2)} \, 
\Biggl[\frac{- \alpha}{(q^2-q^2\hat{\omega}^2 +\alpha)} (v^2-
\hat{\omega}^2-{(\tilde{\bf n}\cdot{\bf v})^2\over\tilde{n}^2}) \nonumber \\
&& \hspace{5.5cm} + \frac{q^2 \mathcal{A}+\mathcal{B}}{q^4 \mathcal{C}+q^2 \mathcal{D}+ \mathcal{E}}\Biggr],
\label{elossReg2}
\eqa
\checked{m}
where 
\bqa
\mathcal{A}&=&(1-\hat{\omega}^2)^2 \beta+\hat{\omega}^2%
{(\tilde{\bf n}\cdot{\bf v})^2\over\tilde{n}^2} (\alpha+\gamma)-2 \hat{%
\omega}(1-\hat{\omega}^2)(\tilde{\bf n}\cdot{\bf v}) \hat{\delta} \nonumber \; ,\\
\mathcal{B}&=&\left((\alpha+\gamma)\beta-\tilde{\bf{n}}^2\hat{\delta}^2\right)%
(1-\hat{\omega}^2-{(\tilde{\bf n}\cdot{\bf v})^2\over\tilde{n}^2})\nonumber \; ,\\
\mathcal{C}&=&-\hat{\omega}^2(1-\hat{\omega}^2)\nonumber \; ,\\
\mathcal{D}&=&-\hat{\omega}^2(\alpha+\gamma)+(1-\hat{\omega}^2)\beta \nonumber%
\\
\mathcal{E}&=&(\alpha+\gamma)\beta-\tilde{\bf{n}}^2\hat{\delta}^2 \, ,
\eqa
\checked{m}
with $\hat{\omega}=\hat{\bf q}\cdot{\bf v}$ and $\hat{\delta}=q \delta$. With all the
momentum dependence made explicit we can now perform the $q$ integration to obtain
\bqa
-\left({{\rm d} W\over {\rm d} x}\right)_{\rm soft} &=& \frac{e^2}{v}  \, {\rm Im} \, %
\int {d \Omega_{q}\over (2 \pi)^3} \, %
\frac{\hat{\omega}}{(1-\hat{\omega}^2)} \, %
\Biggl[-\alpha %
\frac{(v^2-\hat{\omega}^2-{(\tilde{\bf n}\cdot{\bf v})^2\over\tilde{n}^2})}
{2(1-\hat{\omega}^2)}%
\ln{\frac{q^{* 2}(1-\hat{\omega}^2)+\alpha}{\alpha}} \nonumber \\
&& \hspace{5.5cm} + F(q^{\star})-F(0)\Biggr] \,,
\label{Elosssoftfinal}
\eqa
where
\beq
F(q)=\frac{\mathcal{A}}{4 \mathcal{C}} \ln{\left(%
-4\mathcal{C}\left(\mathcal{C} q^4+\mathcal{D}q^2+\mathcal{E}\right)%
\right)}
+\frac{\mathcal{AD}-2\mathcal{BC}}{4 %
\mathcal{C}\sqrt{\mathcal{D}^2-4 \mathcal{CE}}}\ln{\frac{%
\sqrt{\mathcal{D}^2-4 \mathcal{CE}}+\mathcal{D}+2\mathcal{C}q^2}{
\sqrt{\mathcal{D}^2-4 \mathcal{CE}}-\mathcal{D}-2\mathcal{C}q^2}} \; ,
\eeq
and we introduced a UV momentum cutoff $q^{*}$ on the $q$ integration. 

\subsubsection{Small-\protect$\xi$ Limit}

Using a linear expansion in the limit of small $\xi$ one can obtain
analytic expressions for the structure functions $\alpha$, $\beta$, $\gamma$, and $\delta$
as given in Ref.~\cite{RS:2003}. A subsequent expansion of 
Eq.~(\ref{Elosssoftfinal}) to linear order in $\xi$ gives
\beq
-\left({{\rm d} W\over {\rm d} x}\right)_{\rm soft,\,small-\xi} = 
-\left[\left({{\rm d} W\over {\rm d} x}\right)_{\rm soft,iso}+\xi \left(
\left({{\rm d} W\over {\rm d} x}\right)_{\rm soft,\xi_{1}}+%
\frac{({\bf v} \cdot {\bf n})^2}{v^2}%
\left({{\rm d} W\over {\rm d} x}\right)_{\rm soft,\xi_{2}}\right)\right],
\label{softsmallxi}
\eeq
where 
\beq
-\left({{\rm d} W\over {\rm d} x}\right)_{\rm soft,iso} = %
\frac{e^2}{v^2}  \, {\rm Im} \, %
\int_{-v}^{v} {d \hat{\omega} \over (2 \pi)^2} \, %
\hat{\omega}\left[- \frac{v^2-\hat{\omega}^2}{2 (1-\hat{\omega}^2)^2} %
\Pi_T \ln{\frac{(1-\hat{\omega}^2)q^{*2}+\Pi_T}{\Pi_T}}-\frac{\Pi_L}{2}%
\ln{\frac{q^{*2}-\Pi_L}{-\Pi_L}}
\right]
\label{isosoft}
\eeq
\checked{p}
is the isotropic result, which -- assuming $q^{*}\gg m_{D}$ -- 
corresponds to the result from Braaten and Thoma when inserting
the explicit form of $\Pi_T(\hat{\omega})$ and $\Pi_L(\hat{\omega})$ 
from Eqs.~(\ref{pit}) and (\ref{pil}). The form of the 
contributions $\left({{\rm d} W/{\rm d} x}\right)_{\rm soft,\xi_{1}}$
and $\left({{\rm d} W / {\rm d} x}\right)_{\rm soft,\xi_{2}}$ resembles
that of $\left({{\rm d} W / {\rm d} x}\right)_{\rm soft,iso}$, but
since they consist of many more terms than the isotropic contribution we
refrain from giving them here explicitly. 

Note that by scaling out the Debye mass 
$m_D$ from the self-energy functions, the contributions  
$\left({{\rm d} W / {\rm d} x}\right)_{\rm soft,iso}$,
$\left({{\rm d} W /  {\rm d} x}\right)_{\rm soft,\xi_{1}}$ and
$\left({{\rm d} W /  {\rm d} x}\right)_{\rm soft,\xi_{2}}$
depend on
the particle velocity $v$, the ratio $q^{*}/m_D$, and 
an overall multiplicative factor of $e^2 m_D^2$ only. 
The direction of the heavy fermion enters the 
whole result
Eq.~(\ref{softsmallxi}) only through the explicit term 
$({\bf v} \cdot {\bf n})^2/v^2$ in the small $\xi$ limit.
Note that, in principle, the presence of unstable modes
would make the $q$-integration divergent; however, this
is not the case for the soft energy loss as we will discuss in 
Section \ref{sect:shielding}.

\subsubsection{Behavior of the Soft Part}
\label{softsmallxibehavior}

\begin{figure}
\begin{minipage}[t]{.48\linewidth}
\includegraphics[width=0.8\linewidth]{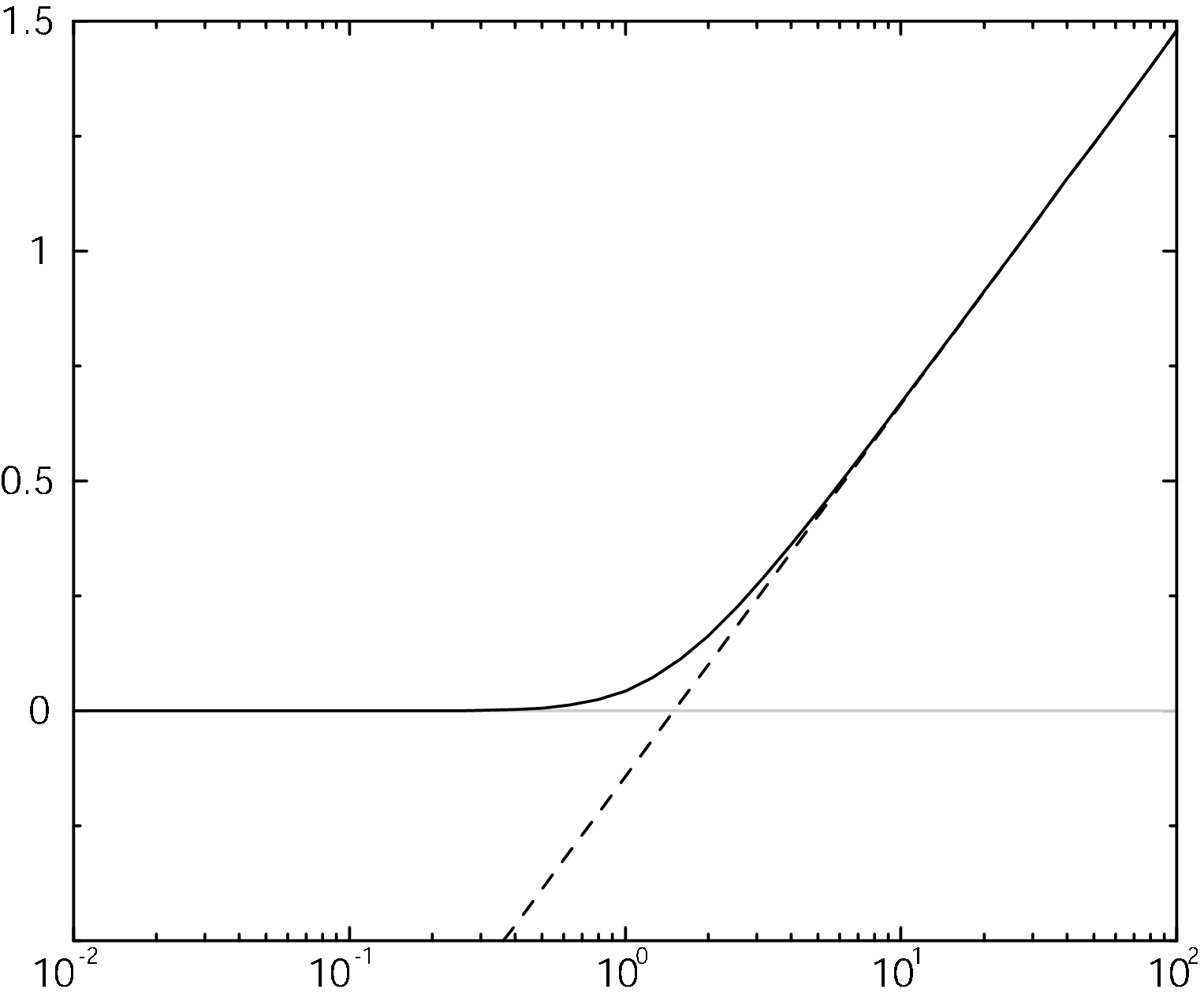}
\setlength{\unitlength}{1cm}
\begin{picture}(9,0)
\put(0.5,1.5){\makebox(0,0){\begin{rotate}{90}%
\footnotesize
$-\left(\frac{dW}{dx}\right)_{\rm soft,iso}%
/\left(\frac{e^4 T^2}{24 \pi}\right)$ 
\end{rotate}}}
\put(4,0.25){\makebox(0,0){\footnotesize $q^{*}/m_D$}}
\end{picture}
\vspace{-1cm}
\caption{Isotropic energy loss of the soft part (full line) as a function
of $q^{*}/m_D$ for $v=0.5$ compared to the result from Braaten and Thoma (dashed line).}
\label{fig:softisoBT}
\end{minipage} \hfill
\begin{minipage}[t]{.48\linewidth}
\includegraphics[width=0.8\linewidth]{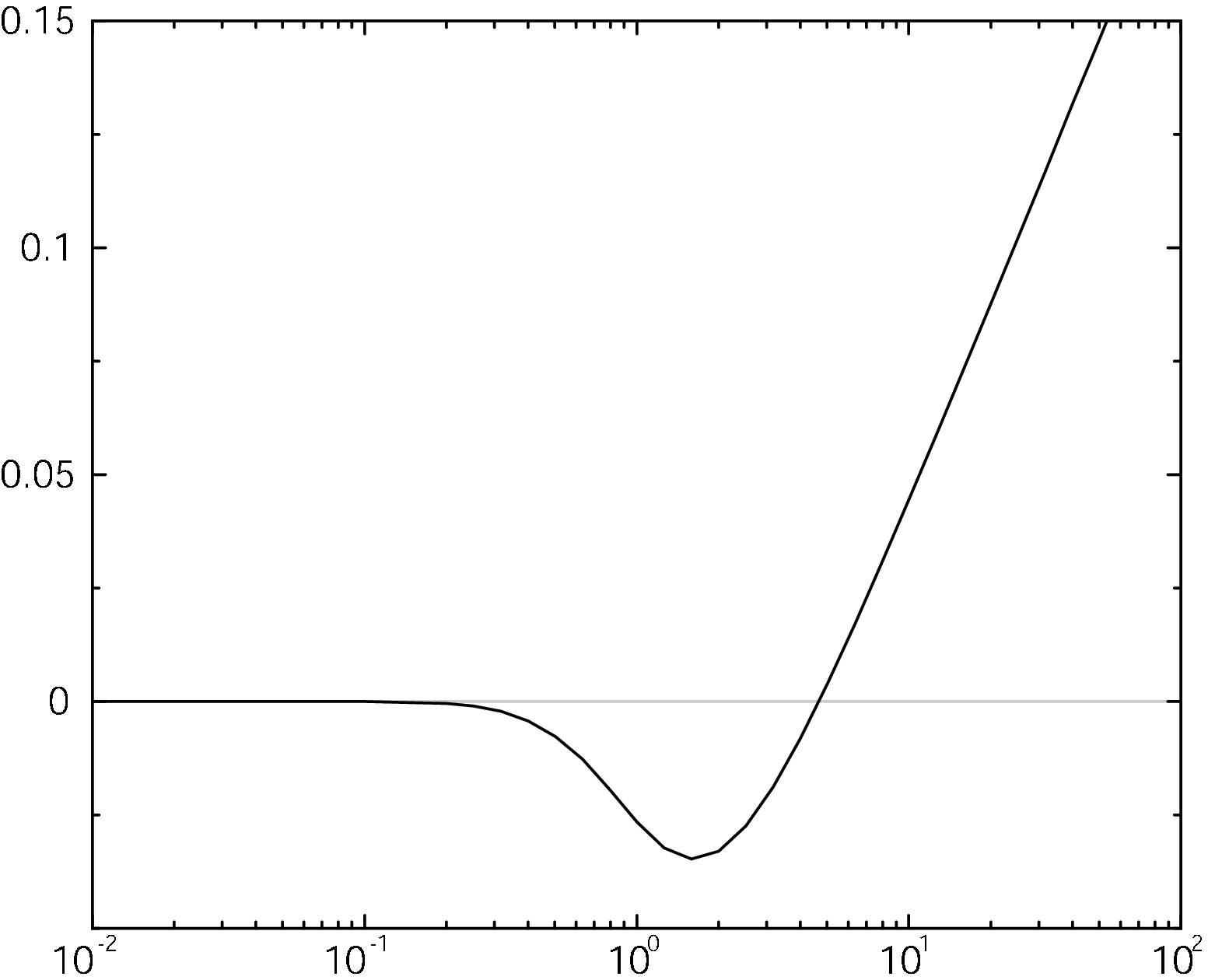}
\setlength{\unitlength}{1cm}
\begin{picture}(9,0)
\put(0.5,1.5){\makebox(0,0){\begin{rotate}{90}%
\footnotesize $-\left(\frac{dW}{dx}\right)_{\rm soft,\xi_2}%
/\left(\frac{e^4 T^2}{24 \pi}\right)$ 
\end{rotate}}}
\put(4,0.25){\makebox(0,0){\footnotesize $q^{*}/m_D$}}
\end{picture}
\vspace{-1cm}
\caption{The contribution $-\left({dW/dx}\right)_{\rm soft,\xi_2}$
as a function of $q^{*}/m_D$ for $v=0.5$.}
\label{fig:softxi2}
\end{minipage} \hfill

\end{figure}

The isotropic result for the soft part Eq.~(\ref{isosoft}) (corresponding to the 
one obtained in Ref.~\cite{GT:1990}) is shown in Figure \ref{fig:softisoBT} 
together with the Braaten-Thoma result from Ref.~\cite{BT:1991} for $v=0.5$.  As 
can be seen, the results are identical for large $q^{*}/m_D$, while for small 
$q^{*}/m_D$ the Braaten-Thoma result becomes negative whereas the full result 
obtained by numerical integration of Eq.~(\ref{isosoft}) is positive for all 
values of $q^{*}/m_D$.

In Figure \ref{fig:softxi2} we show the function $-\left({{\rm d} W / {\rm d} 
x}\right)_{\rm soft,\xi_{2}}$ which controls the directional dependence of the 
soft part of the energy loss for small $\xi$. As can be seen from this Figure 
the function is negative for small $q^*/m_D$ but becomes positive at a velocity-dependent 
value of $q^*/m_D$.  The value of $q^{*}/m_D$ where $\left({{\rm d} 
W / {\rm d} x}\right)_{\rm soft,\xi_{2}}=0$ is of special interest here since it 
signals the point at which the directional dependence of the energy loss 
changes.  The value of $q^*/m_D$ obtained using the principle
of minimal sensitivity is a function of the velocity and coupling constant
and therefore we expect a non-trivial dependence on these parameters.  
For $\xi>0$ this means that as the coupling constant is increased from zero 
for fixed $v$ that at first the energy loss will be peaked along ${\bf n}$ but 
will eventually become peaked transverse to ${\bf n}$.  The value of the 
coupling at which this change in the directional dependence occurs increases as 
the velocity of the fermion increases. For the exact expression given by 
Eq.~(\ref{Elosssoftfinal}) the results obtained in the small $\xi$ limit hold 
qualitatively but quantitatively the predictions differ considerably once $\xi$ 
becomes large.

\subsubsection{Large-\protect$\xi$ Limit}

In the limit $\xi\rightarrow \infty$ one finds that the anisotropic 
distribution function turns into a specific case of Eq.(\ref{deltalimit}),
\beq
\lim_{\xi \rightarrow \infty} \sqrt{1+\xi} \; n_F\left(p\sqrt{1+\xi(%
{\bf \hat{p}}\cdot {\bf \hat{n}})^2}\right)=%
\delta ({\bf \hat{p}}\cdot {\bf \hat{n}}) \int_{-\infty}^{\infty} dx
\; n_F\left(p\sqrt{1+x^2}\right) \; ,
\label{largexilimit}
\eeq
where the remaining integral representing $F(p_{\perp})$ in 
Eq.(\ref{deltalimit}) can be evaluated analytically
to be a sum over Bessel functions.
Following Ref.~\cite{ALM:2003} it is then possible to evaluate the structure
functions analytically. The result for the soft part of the energy loss
is then given by Eq.(\ref{Elosssoftfinal}) with the general structure
functions replaced by their large-$\xi$ expressions.
For large values of $\xi$, the general result for the soft part
converges towards the limiting result, as it should.

\subsection{Dynamical Shielding of Plasma Instabilities}
\label{sect:shielding}

As shown previously \cite{Weibel:1959,SM:1993,SM:1994,SM:1997,RM:2003,RS:2003,ALM:2003}
in systems in which the distribution function is anisotropic in momentum space there exist
unstable modes which could potentially make physical quantities incalculable
within a perturbative framework~\cite{M:2003}.  However, we will show in this section that 
in the case of the energy loss the singularities induced by the presence of these plasma 
instabilities are ``shielded'' and are therefore rendered safe.  To see this one need only 
consider the contribution from the $\Delta_A$ propagator to Eq.~(\ref{elossReg2}) which is 
schematically given by
\bqa
\left({{\rm d} W\over {\rm d} x} \right)_{\rm A,\,soft} &\sim& 
{\rm Im} \, \int d\Omega \int q\,dq \, \frac{\hat{\omega}}{(1-\hat{\omega}^2)} \, 
\frac{- \alpha}{(q^2-q^2\hat{\omega}^2 +\alpha)} \; .
\label{elossSchematic}
\eqa
The possibility of singularities arise because in the static limit the structure
function $\alpha$ is negative-valued which can result in singularities along the integration
path which are unregulated.  However, in the limit $\omega\rightarrow0$ the structure 
function $\alpha$ has the form
\bqa
\lim_{\omega\rightarrow0} \alpha(\omega,q) = M^2(-1 + i D \hat\omega) + {\cal O}(\omega^2) \; ,
\eqa
where $M$ and $D$ depend on the angle of propagation with respect to the anisotropy
vector and the strength of the anisotropy.  

Ignoring the ${\cal O}(\omega)$ contribution above we see that a singularity in the integrand can arise 
at the point $q=M$ when $\hat\omega \rightarrow 0$.  However, including the 
${\cal O}(\omega)$ contribution we see that the combination of the power of $\hat\omega$ in
the numerator and the presence of a contribution proportional to $\hat\omega$ in 
$\alpha$ together render the singularity finite at this point as long as $D$ is non-vanishing
\bqa
\lim_{\hat\omega\rightarrow 0} \, \lim_{q\rightarrow M} 
\frac{\hat\omega}{q^2-q^2\hat{\omega}^2 + M^2(-1+iD\hat\omega)} 
\;\rightarrow\; \frac{1}{M^2(\hat{\omega}+ i D)} 
\;\rightarrow\; -\frac{i}{M^2 D} 
\; .
\label{elossSchematic2}
\eqa
Therefore, the $q$-integration can safely be performed.  Of course, the integrand
may change rapidly in this region so it should be treated carefully when numerically
evaluating the integral.  Note that this is similar to dynamical screening 
of the magnetic sector of finite temperature QCD.  In fact, in the isotropic limit, 
the coefficient $D$ approaches 
the well-known value of $-i\pi/4$ which is due to Landau damping.
Unfortunately for anisotropic systems it is difficult to prove that the coefficient $D$
is non-vanishing for all values of the propagation angle and anisotropy strength.
However, we can evaluate $D$ analytically in the weak-anisotropy limit and 
numerically for general $\xi$ and in both cases we find that $D$ is non-vanishing.

\begin{figure}
\begin{minipage}[t]{.48\linewidth}
\includegraphics[width=0.8\linewidth]{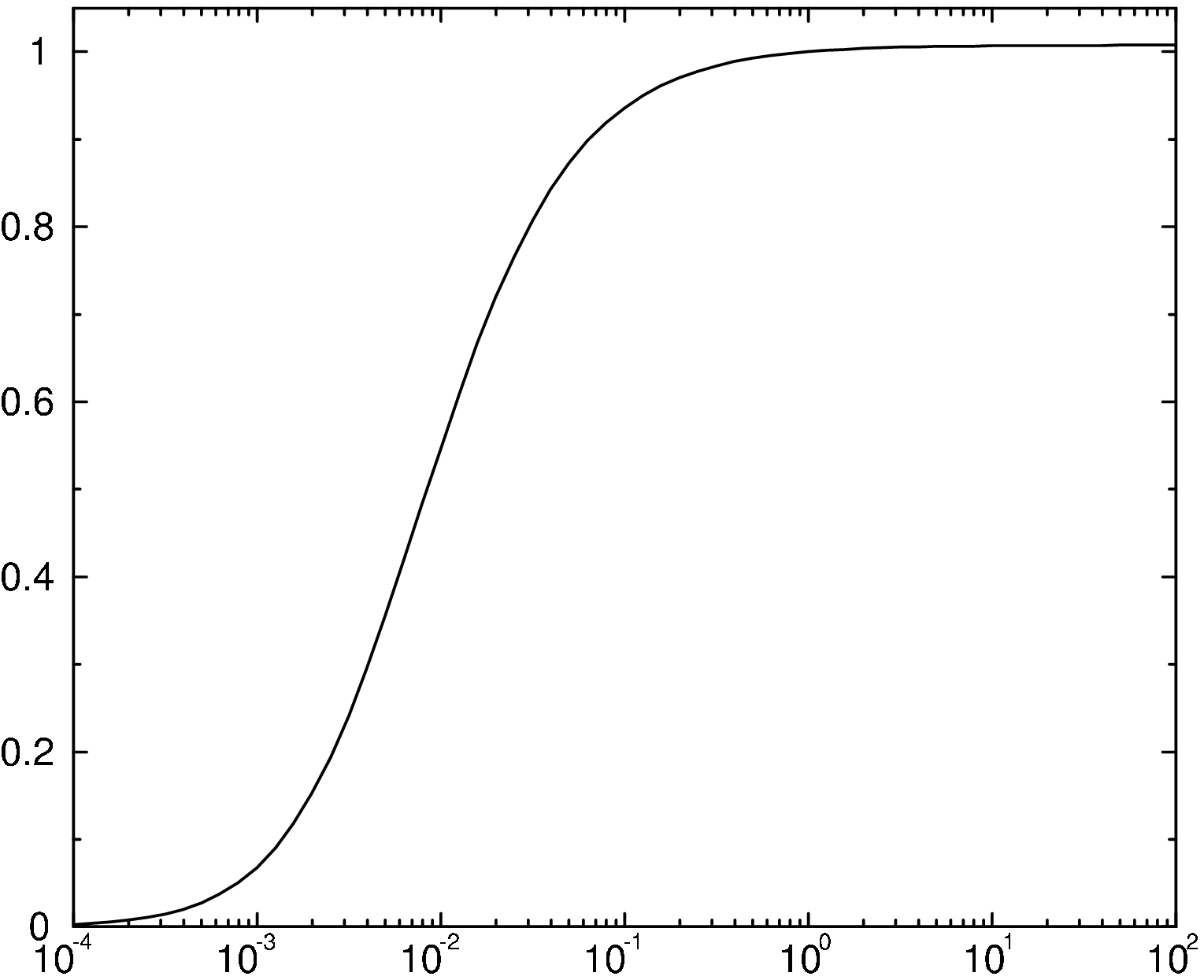}
\setlength{\unitlength}{1cm}
\begin{picture}(9,0)
\put(0.5,2.3){\makebox(0,0){\begin{rotate}{90}%
\footnotesize
$- 4 D / \pi$
\end{rotate}}}
\put(4,0.2){\makebox(0,0){\footnotesize $1+\xi$}}
\end{picture}
\vspace{-1cm}
\caption{
Coefficient $D$ as a function of $\xi$ for $\theta_n=1.5$.
}
\label{fig:Dfig1}
\end{minipage} \hfill
\begin{minipage}[t]{.48\linewidth}
\includegraphics[width=0.8\linewidth]{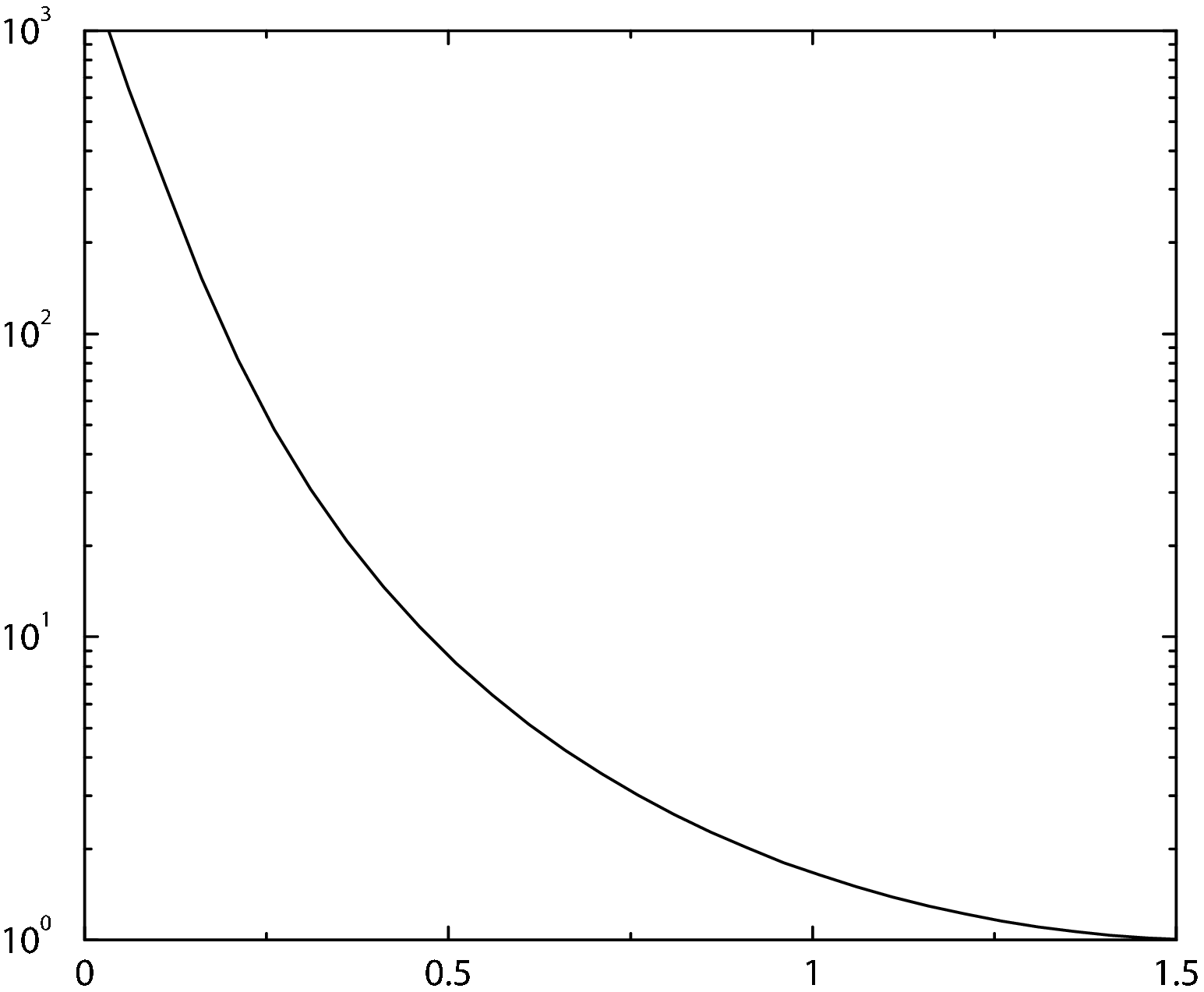}
\setlength{\unitlength}{1cm}
\begin{picture}(9,0)
\put(0.5,2.3){\makebox(0,0){\begin{rotate}{90}%
\footnotesize 
$- 4 D / \pi$
\end{rotate}}}
\put(4,0.2){\makebox(0,0){\footnotesize $\theta_n$}}
\end{picture}
\vspace{-1cm}
\caption{
Coefficient $D$ as a function of $\theta_n$ for $\xi=100$.
}
\label{fig:Dfig2}
\end{minipage} \hfill
\end{figure}

In the weak-anisotropy limit we can derive analytic expressions for 
all of the structure functions as shown in Ref.~\cite{RS:2003}.
To linear order in $\xi$ the structure function $\alpha$ becomes
\bqa
\lim_{\hat\omega\rightarrow0} \, \lim_{\xi\rightarrow0} 
\alpha = m_D^2 \left[ - {\xi\over6} (1+\cos2\theta_n) 
- {i \pi \over 4} \hat\omega \left( 1
 + {3 \xi \over 4} (1+\cos2\theta_n) 
 \right) \right] \, ,
\eqa
where $\theta_n$ is the angle of propagation with respect to the anisotropy vector
${\bf n}$.  Therefore, in the weak-anisotropy limit $D$ is non-vanishing for all 
$\xi>0$.  For $\xi<0$, the $\Delta_A$ mode is stable so there is no singularity to be 
concerned about.  At higher orders in the weak-anisotropy 
expansion the picture becomes slightly more complicated but is still 
consistent with $D$ being non-vanishing.  

For large values of $\xi$ we can not rely on a weak-anisotropy expansion so we have to resort 
to numerical determination of $D$.  In Fig.~\ref{fig:Dfig1} we plot the dependence 
of $D$ on $\xi$ for $\theta_n=1.5$ and in Fig.~\ref{fig:Dfig2} we plot the dependence 
of $D$ on $\theta_n$ for $\xi=100$.  As can be seen from these Figures, $D$ is 
non-vanishing for all values of $\xi$ and $\theta_n$ shown.  Beyond what we have plotted,
we have calculated $D$ for numerous values of $\xi$ and $\theta_n$ and in every case we
find that $D$ is non-vanishing.  We are therefore reasonably confident
that $D$ is in general a non-vanishing quantity and therefore the singularities 
which could come from the unstable modes are dynamically shielded and thereby 
rendered safe in the energy loss calculation.
Note that similar arguments can be made to show that the singularities coming from the 
$\Delta_G$ contribution to Eq.~(\ref{elossReg2}) are also dynamically shielded.

\subsection{Hard Part}
\label{hardpart}
\begin{figure}[t]
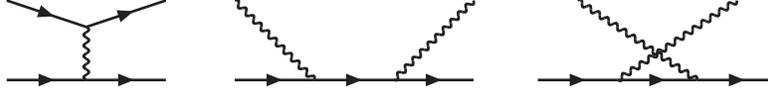

\bqa \nonumber
\picb{\Lqu(0,0)(30,0) \Lqu(30,0)(60,0) \Lgl(30,20)(30,0) \Lqu(0,30)(30,20) \Lqu(30,20)(60,30)}~
\hspace{1cm}
\picb{\Lqu(0,0)(30,0) \Lqu(30,0)(60,0) \Lqu(60,0)(90,0) \Lgl(0,30)(30,0) \Lgl(90,30)(60,0)}~
\hspace{2cm}
\picb{\Lqu(0,0)(30,0) \Lqu(30,0)(60,0) \Lqu(60,0)(90,0) \Lgl(75,30)(30,0) \Lgl(15,30)(60,0)}
\hspace{1.75cm}
\eqa
\caption[a]{Tree-level Feynman diagrams for the scattering processes $e^-\mu\rightarrow e^-\mu$ and 
$\gamma \mu \rightarrow \gamma \mu$. Note that the second two diagrams cancel with each other so
that in QED only the first diagram contributes to the hard energy loss.}
\label{hard-diagrams}
\end{figure}

The  expression for the hard contribution to the energy loss for an arbitrary
electron distribution function $f_e({\bf p})$ is obtained by summing the tree level diagrams
shown if Figure \ref{hard-diagrams}.   Assuming $v \gg T/E$, performing the Dirac traces, and
summing over spins the result can be reduced to \cite{BT:1991}
\bqa
-\left({\rm d} W \over {\rm d} x\right)_{\rm hard} = {4 \pi e^4 \over v}
	\int {d^3{\bf k} \over (2 \pi)^3} {f_e({\bf k})\over k} 
	\int {d^3{\bf k^\prime} \over (2 \pi)^3} {1 - f_e({\bf k}^\prime) \over k^\prime}
	\delta(\omega-{\bf v}\cdot{\bf q}) \Theta(q-q^*) \nonumber \\
	\times \; {\omega \over (\omega^2-q^2)^2} 
	\left[ 2 (k-{\bf v}\cdot{\bf k}) (k^\prime-{\bf v}\cdot{\bf k}^\prime)
		+ {1 - v^2 \over 2} (\omega^2 - q^2) \right] \; ,
\eqa
\checked{mp}
where $\omega = k^\prime - k$ and ${\bf q} = {\bf k}^\prime - {\bf k}$, and we 
have introduced an infrared cutoff $q^*$ on the $q$ integration.  The term 
involving the product $f_e({\bf k}) f_e({\bf k}^\prime)$ vanishes since the 
integrand is odd under the interchange of ${\bf k}$ and ${\bf k}^\prime$. 
Redefining the origin of the ${\bf k}^\prime$
integration so that it becomes an integration over ${\bf q}$ we have
\bqa
-\left({\rm d} W \over {\rm d} x\right)_{\rm hard} &=& {e^4 \over 2 \pi^2 v}
	\int {d^3{\bf k} \over (2 \pi)^3} {f_e({\bf k})\over k} 
	\int_{q^*}^\infty q^2 dq \, \int d\Omega_{q} \;
	\frac{\delta(\omega-{\bf v}\cdot{\bf q})}{|\bf{q}+\bf{k}|} \nonumber \\
	\hspace{1.5cm} && \times \; {\omega \over (\omega^2-q^2)^2} 
	\left[ 2 (k-{\bf v}\cdot{\bf k}) (\omega+k- {\bf v}\cdot{\bf k}- {\bf v}\cdot{\bf q})
		+ {1 - v^2 \over 2} (\omega^2 - q^2) \right] \; ,
\label{elosshard1}
\eqa
\checked{mp}
where now $\omega=|{\bf q}+{\bf k}|-k.$ Choosing $\bf{v}$ to be the z-axis for the 
$\bf{q}$ and $\bf{k}$ integration
we can rewrite the delta-function as 
\beq
\delta(|{\bf q}+{\bf k}|-k-{\bf v}\cdot{\bf q})=\frac{\delta(\phi_q-\phi_{0})%
\Theta(k+{\bf v}\cdot{\bf q}) 2 |{\bf q}+{\bf k}|}
{q \sqrt{4 k^2 \sin^2 \theta_k \sin^2 \theta_q-(q(1-v^2 \cos^2 \theta_q)+%
2 \cos \theta_q (k \cos \theta_k-k v))^2}},
\eeq
\checked{mp}
where $\phi_{0}$ is the solution of the equation 
\beq
\cos (\phi_0-\phi_k)=-\frac{q(1-v^2 \cos^2 \theta_q)+2 \cos \theta_q (k \cos %
\theta_k-k v)}{2 k \sin \theta_k \sin \theta_q}.
\eeq
\checked{mp}
%
%
The $\phi_q$ integration is then straightforward and we obtain
\bqa
-\left({\rm d} W \over {\rm d} x\right)_{\rm hard} &=& {e^4 \over 2 \pi^2 v}
	\int {d^3{\bf k} \over (2 \pi)^3} {f_e({\bf k})\over k} 
	\int_{q^*}^\infty q \, dq \, \int_{-1}^{1} d\cos{\theta_{q}} \;
	4 \Theta(k+v q \cos{\theta_q})\nonumber \\
	&& \hspace{3mm} \times \;
	\frac{\Theta(4 k^2 \sin^2 \theta_k \sin^2 \theta_q-(q(1-v^2 \cos^2 \theta_q)+%
	2 \cos \theta_q (k \cos \theta_k-k v))^2)}%
	{\sqrt{4 k^2 \sin^2 \theta_k \sin^2 \theta_q-(q(1-v^2 \cos^2 \theta_q)+%
	2 \cos \theta_q (k \cos \theta_k-k v))^2}}
	\nonumber \\
	&& \hspace{7mm} \times \; {\omega \over (\omega^2-q^2)^2} 
	\left[ 2 (k-{\bf v}\cdot{\bf k})^2
		+ {1 - v^2 \over 2} (\omega^2 - q^2) \right] \; ,
\label{myelosshard2}
\eqa
\checked{mp}
where $\omega=v q \cos{\theta_q}$ and we 
have included a factor of $2$ because of the symmetry $\phi_{0} \leftrightarrow 2\pi-\phi_{0}$.  
Setting $f_{e,{\rm iso}}(k) = n_F(k)$ and scaling $k=q z$ one can perform the $q$ integration to obtain
\bqa
-\left({\rm d} W \over {\rm d} x\right)_{\rm hard} &\!\!\!=\!\!& %
{8 e^4 (\hat{q}^{*})^2 T^2 \sqrt{1+\xi} \over (2 \pi)^5 v}
	\int_{0}^{\infty} z dz \int_{-1}^{1} d\cos \theta_k 
	\nonumber \\
	&& \hspace{-23mm} \times \;
	\left[
	\int_{0}^{2 \pi} \! d \phi_k F_1 (\hat{q}^* z
 	\sqrt{1+\xi (n_x \sin \theta_k
	\cos \phi_k+n_z \cos \theta_k)^2})\right] \nonumber \\
	&& \hspace{-23mm} \times \;
	\left\{
	\int_{-1}^{1} d\cos{\theta_{q}} \;
	\Theta(z+v \cos{\theta_q})
	\frac{\Theta(4 z^2 \sin^2 \theta_k \sin^2 \theta_q-(1-v^2 \cos^2 \theta_q+%
	2 \cos \theta_q z (\cos \theta_k- v))^2)}%
	{\sqrt{4 z^2 \sin^2 \theta_k \sin^2 \theta_q-(1-v^2 \cos^2 \theta_q+%
	2 \cos \theta_q z (\cos \theta_k- v))^2}} \right. \nonumber \\
	&& \hspace{-13mm} \left.	\times \;
	{v \cos \theta_q \over (v^2 \cos^2 \theta_q-1)^2} 
	\left[ 2 z^2(1-v \cos \theta_k)^2
		+ {1 - v^2 \over 2} (v^2 \cos^2 \theta_q - 1) \right]%
 \;\right\} \; ,
\label{myelosshard3}
\eqa
\checked{p}
where $\hat{q}^{*}=q^{*}/T$,
\beq
F_1(x)=\frac{x \ln{(1+\exp{(-x)})} -{\rm Li}_2(-\exp{(-x)})}{x^2} \, ,
\label{nastyfunction}
\eeq
\checked{p}
and $n_z=\bf{n}\cdot\hat{\bf v}$ with $1=n_x^2+n_z^2$. To obtain the final 
result for the hard contribution to the energy loss the remaining integrations 
have to be performed numerically.

\subsubsection{Small-\protect$\xi$ Limit}

For small $\xi$ we use Eq.~(\ref{elosshard1}) and scale out the 
$\xi$-dependence using Eq.~(\ref{distfunc2})
and substituting $k^2\rightarrow k^2(1+\xi (\hat{{\bf k}}\cdot {\bf n})^2)$
and $q^2\rightarrow q^2(1+\xi (\hat{{\bf k}}\cdot {\bf n})^2)$.
Performing a linear expansion in $\xi$ then gives
\beq
-\left({\rm d} W \over {\rm d} x\right)_{\rm hard,\,small-\xi} = 
{e^4 \over 2 \pi^2 v}
\int_0^{\infty} dk\, f_{e,{\rm iso}}(k) \,\left(%
\int_{q^*}^\infty dq \, \left[g_1(q)-\xi \left(g_2(q)-{g_1(q)\over2}\right) \right]
-\frac{\xi}{2} q^{*} g_{2}(q^*)
\right),
\eeq
where

\bqa
g_{1}(q)&=& \int d {\Omega_q} \int \frac{d {\Omega_k}}{4\pi} q^2 \frac{%
\delta(\omega-{\bf v}\cdot{\bf q})}{|{\bf q}+{\bf k}|} \frac{\omega}{\omega^2-%
q^2}\left[2 (k-{\bf v}\cdot{\bf k})^2+\frac{1-v^2}{2}(\omega^2-q^2)\right]%
\nonumber \\
g_{2}(q)&=& \int d {\Omega_q} \int \frac{d {\Omega_k}}{4\pi} q^2 %
({\bf \hat{k}}\cdot {\bf n})^2 \frac{%
\delta(\omega-{\bf v}\cdot{\bf q})}{|{\bf q}+{\bf k}|} \frac{\omega}{\omega^2-%
q^2}\left[2 (k-{\bf v}\cdot{\bf k})^2+\frac{1-v^2}{2}(\omega^2-q^2)\right].
\label{f1f2}
\eqa

Taking ${\bf q}$ as the z-direction makes the $d \Omega_k$ integration 
straightforward but still algebraically intensive for the $g_2$ contribution because 
of the extra dependence on ${\bf \hat{k}}\cdot {\bf n}$. Next we rotate the 
coordinate system so that ${\bf v}$ becomes the new z-direction and ${\bf n}$ 
lies in the $x-z$ plane for the $d \Omega_q$ integration; the polar integration 
is then also straightforward whereas for the azimuthal integration the limits 
coming from the delta function in Eq.~(\ref{f1f2}) are somewhat non-trivial. 
Nevertheless, by scaling $k=q z$ one can perform the $q$ integration to obtain
\beq
-\left({\rm d} W \over {\rm d} x\right)_{\rm hard,\, small-\xi} \!\! = 
{e^4 (q^*)^2 \over 2 \pi^2 v}
\int_0^{\infty} \frac{dz}{2 \pi^2}\, z \left[
F_1(\hat{q}^* z) \left[ g_1(1) -\xi \left(g_2(1)-{g_1(1)\over2}\right)\right]
-\frac{\xi}{2} f_{\rm iso}(\hat{q}^* z) g_2(1)
\right],
\label{hardsmallxi}
\eeq
where $\hat{q}^{*}=q^{*}/T$, 
\bqa
g_1(1)&=&\frac{\pi}{z v} \int_{|1-z|-z}^{1} d\omega \, \Theta(v^2-\omega^2) \,
\omega \Biggl[ {3 \omega^2 \over 4} - {v^2\over4} + 
3 z (z+\omega) \nonumber \\
	&& \hspace{6cm}  
		- {1-v^2 \over 2} {1 \over 1 -\omega^2}
		- (1-v^2) {z(z+\omega)\over 1 -\omega^2}
		\Biggr] \; ,
\eqa
and the exact form of $g_2$ is similar to $g_1$ but involves many more
terms so we have refrained from giving an explicit expression.
The integration over $\omega$ can then be performed analytically using
\beq
\int_0^{\infty} dz \, \int_{|1-z|-z}^1 d\omega \, \Theta(v^2-\omega^2) = 
  \int_{(1-v)/2}^{(1+v)/2} dz \, \int_{1-2z}^v d \omega
  +
  \int_{(1+v)/2}^{\infty} dz \, \int_{-v}^v d \omega \, ,
\eeq
but the remaining integral over $z$ has to be performed numerically.  
The final result to linear order in $\xi$ can then be written as
\beq
-\left({\rm d} W \over {\rm d} x\right)_{\rm hard, small-\xi} = 
-\left[\left({\rm d} W \over {\rm d} x\right)_{\rm hard,iso}+\xi
\left(\left({\rm d} W \over {\rm d} x\right)_{\rm hard,\xi_1}
+ \frac{({\bf v}\cdot {\bf n})^2}{v^2} 
\left({\rm d} W \over {\rm d} x\right)_{\rm hard,\xi_2}\right)\right],
\eeq
where $\left({\rm d} W / {\rm d} x\right)_{\rm hard,iso}$ is given
by Eq.~(\ref{hardsmallxi}) with $\xi=0$.

\subsubsection{Behavior of the Hard Part}
\label{hardsmallxibehavior}

\begin{figure}
\begin{minipage}[t]{.48\linewidth}
\includegraphics[width=0.8\linewidth]{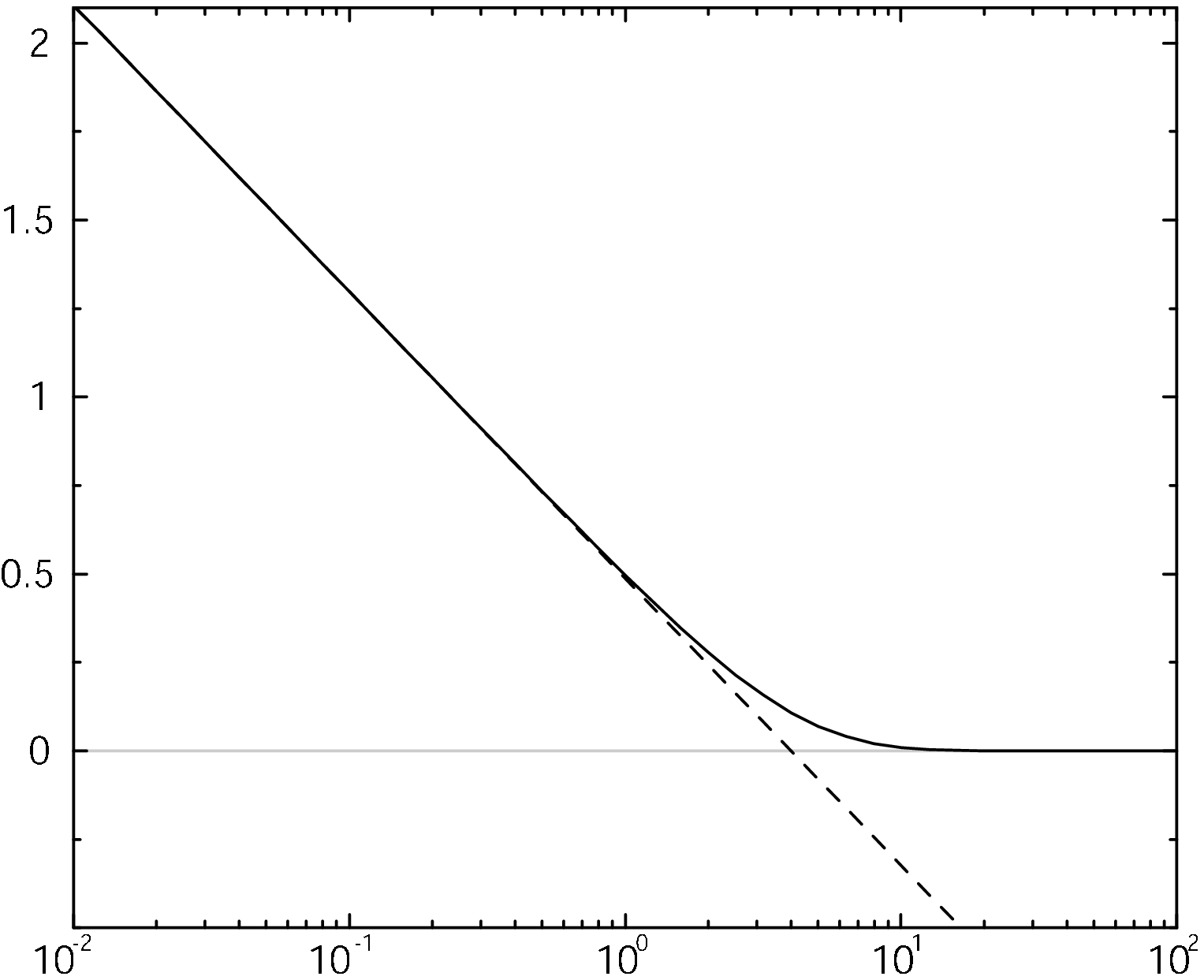}
\setlength{\unitlength}{1cm}
\begin{picture}(9,0)
\put(0.5,1.7){\makebox(0,0){\begin{rotate}{90}%
\footnotesize
$-\left(\frac{dW}{dx}\right)_{\rm hard,iso}%
/\left(\frac{e^4 T^2}{24 \pi}\right)$ 
\end{rotate}}}
\put(4,0.2){\makebox(0,0){\footnotesize $\hat{q}^{*}$}}
\end{picture}
\vspace{-1cm}
\caption{Isotropic energy loss of the hard part (full line) as a function
of $\hat{q}^{*}$ for $v=0.5$ compared to the result from Braaten and Thoma (dashed line).}
\label{fig:hardisoBT}
\end{minipage} \hfill
\begin{minipage}[t]{.48\linewidth}
\includegraphics[width=0.8\linewidth]{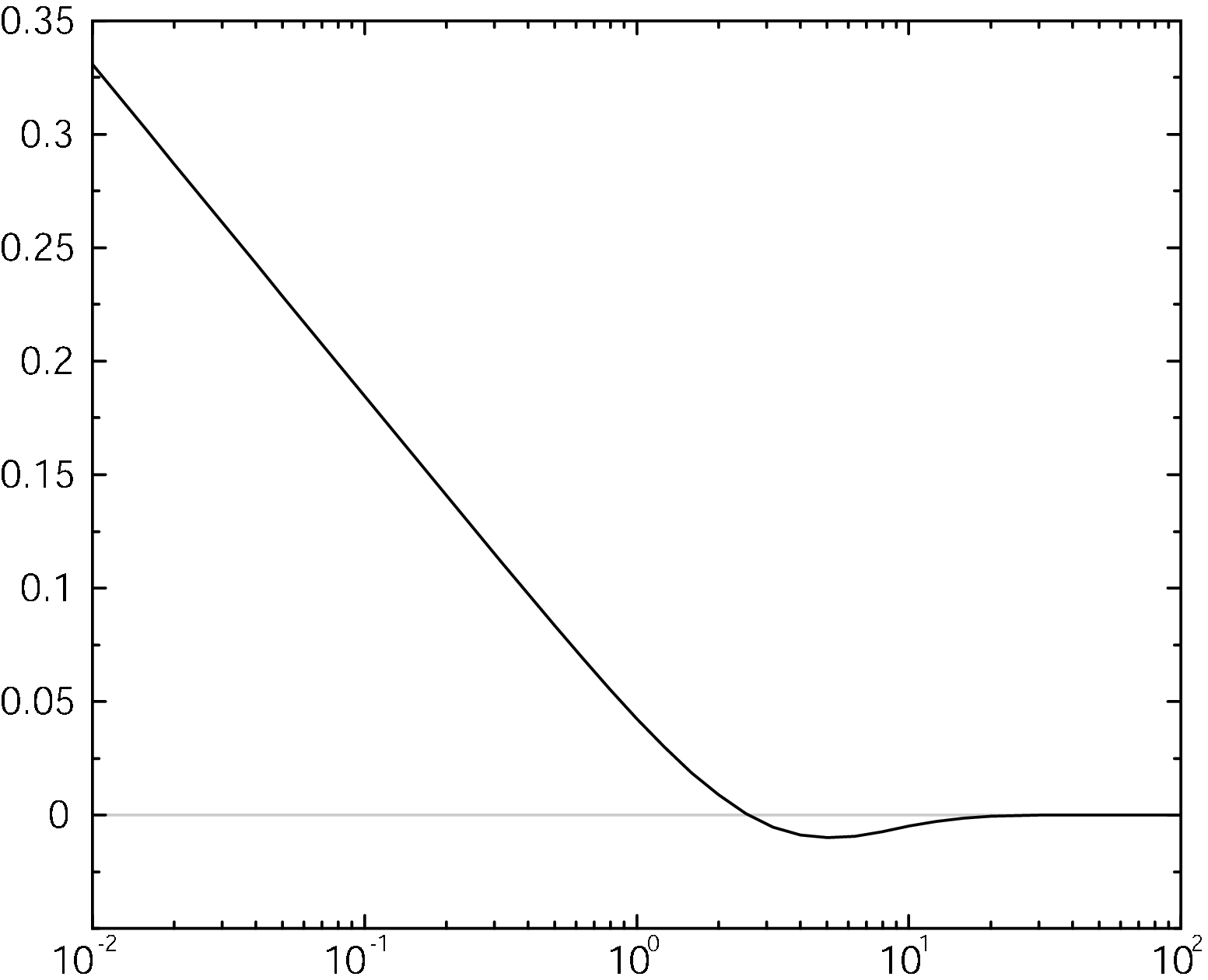}
\setlength{\unitlength}{1cm}
\begin{picture}(9,0)
\put(0.5,1.7){\makebox(0,0){\begin{rotate}{90}%
\footnotesize $-\left(\frac{dW}{dx}\right)_{\rm hard,\xi_2}%
/\left(\frac{e^4 T^2}{24 \pi}\right)$ 
\end{rotate}}}
\put(4,0.2){\makebox(0,0){\footnotesize $\hat{q}^{*}$}}
\end{picture}
\vspace{-1cm}
\caption{The contribution $-\left({dW/dx}\right)_{\rm hard,\xi_2}$
as a function of $\hat{q}^*$ for $v=0.5$.}
\label{fig:hardxi2}
\end{minipage} \hfill
\end{figure}

In Figure \ref{fig:hardisoBT} we show the result for the isotropic ($\xi=0$) 
result for the hard part compared to the Braaten-Thoma result from 
Ref.~\cite{BT:1991}. Similar to what we found for the soft part, the two results 
are identical for very small $\hat{q}^{*}$, while for large $\hat{q}^{*}$ the 
Braaten-Thoma result becomes negative and our result is positive for 
all values of $\hat{q}^{*}$.

In Figure \ref{fig:hardxi2} we plot the function 
$-\left({dW/dx}\right)_{\rm hard,\xi_2}$ which controls the 
directional dependence of the hard part of the energy loss. As was the
case with the soft part, $\left({dW/dx}\right)_{\rm hard,\xi_2}=0$ for 
a velocity-dependent value of $q^{*}$.  Again, in practice, this means
that the directional dependence of the hard energy loss changes as
the coupling constant is increased. 

\subsubsection{Large-\protect$\xi$ Limit}

Using the explicit form (\ref{largexilimit}) for $f_e$ one can do
the $d\phi_k$ integration in Eq.(\ref{myelosshard2}) by rewriting
\beq
\delta({\bf \hat{k}}\cdot {\bf \hat{n}})=\frac{\delta(\phi_k-\phi_0)}{%
\sqrt{n_x^2-\cos^2{\theta_k}}}, \quad %
\cos{\theta_0}=-\frac{\cos{\theta_k} n_z}{\sin{\theta_k} n_x}.
\eeq
After scaling $\cos{\theta_k}\rightarrow n_x \cos{\theta_k}$ and 
doing the $q$-integration the remaining integrals take the form
\bqa
-\left({\rm d} W \over {\rm d} x\right)_{\rm hard} &\!\!\!=\!\!& %
{32 e^4 (\hat{q}^{*})^2 T^2 \over (2 \pi)^5 v}%
	\int_{0}^{\infty} z dz \int_{0}^{\infty} dx%
	F_1 (\hat{q}^* z\sqrt{1+x^2}) \int_{-1}^{1} d\cos \theta_k %
\frac{1}{\sqrt{1-\cos^2{\theta_k}}} \int_{-1}^{1} d\cos{\theta_{q}}
	\nonumber \\
	&& \hspace{-13mm} \left\{
	\Theta(z+v \cos{\theta_q})
	\frac{\Theta(4 z^2 (1- n_x^2\cos^2 \theta_k) \sin^2 \theta_q-(1-v^2 \cos^2 \theta_q+%
	2 \cos \theta_q z (n_x \cos \theta_k- v))^2)}%
	{\sqrt{4 z^2 (1- n_x^2\cos^2 \theta_k) \sin^2 \theta_q-(1-v^2 \cos^2 \theta_q+%
	2 \cos \theta_q z (n_x \cos \theta_k- v))^2}} \right. \nonumber \\
	&& \hspace{-13mm} \left.	\times \;
	{v \cos \theta_q \over (v^2 \cos^2 \theta_q-1)^2} 
	\left[ 2 z^2(1-v n_x \cos \theta_k)^2
		+ {1 - v^2 \over 2} (v^2 \cos^2 \theta_q - 1) \right]%
 \;\right\} \; ,
\label{myelosshardLX}
\eqa
which after evaluation then give the result for the hard
part of the energy loss in the large-$\xi$ limit.

\section{Numerical evaluation}
\label{results}

\subsection{Isotropic Case}

In the isotropic case, the total collisional energy loss is
obtained by adding Eq. (\ref{isosoft}) and (\ref{hardsmallxi})
with $\xi=0$,
\beq
\left({\rm d} W \over {\rm d} x\right)_{\rm iso} = 
\left({\rm d} W \over {\rm d} x\right)_{\rm iso,soft} + 
\left({\rm d} W \over {\rm d} x\right)_{\rm iso,hard} \; .
\eeq
In general the isotropic energy loss is a 
function of the electromagnetic coupling $e$, 
the velocity of the particle $v$, the temperature $T$, and
the momentum separation scale $q^{*}$. 

\begin{figure}
\begin{minipage}[t]{.48\linewidth}
\includegraphics[width=0.8\linewidth]{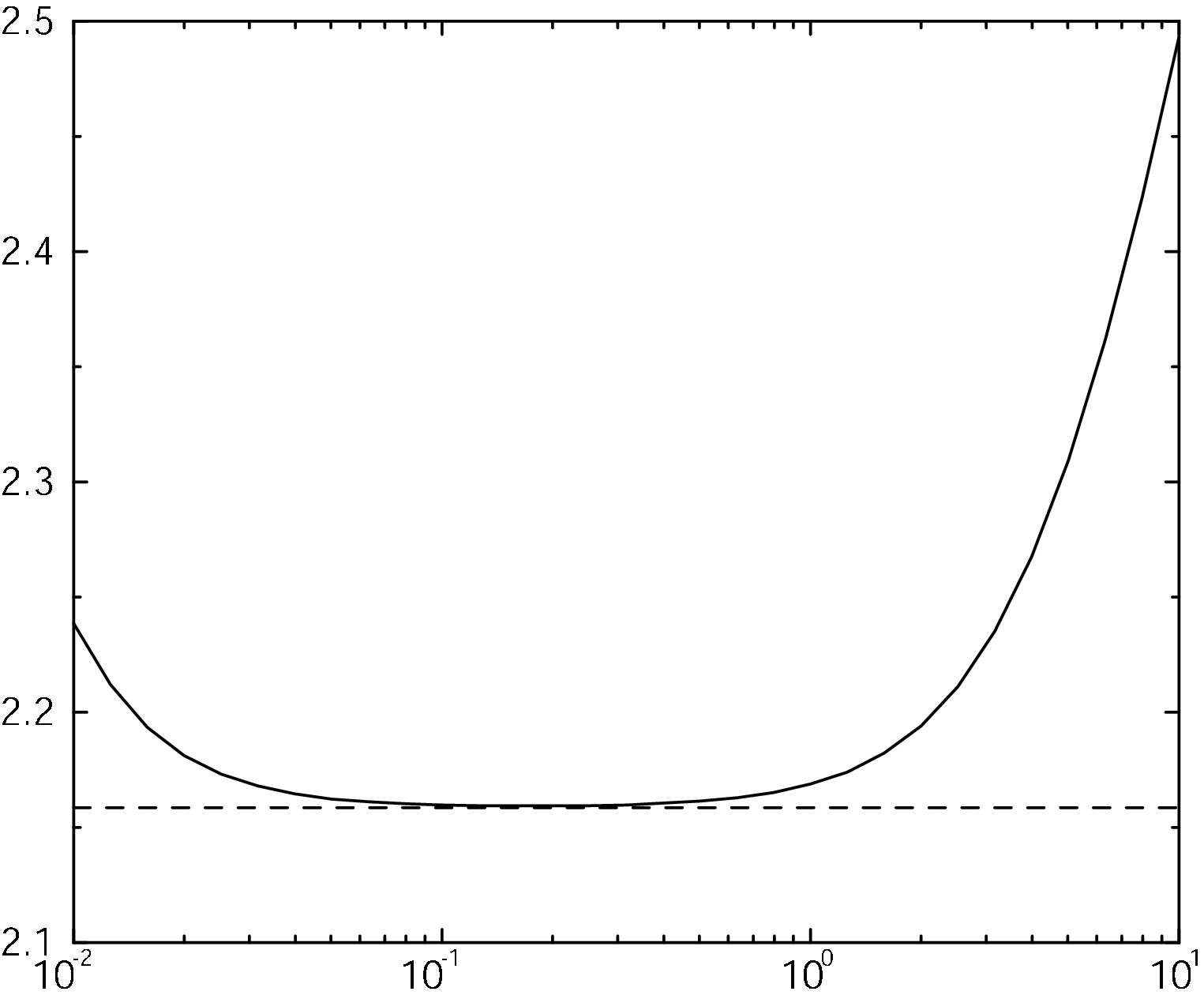}
\setlength{\unitlength}{1cm}
\begin{picture}(9,0)
\put(0.2,2.2){\makebox(0,0){\begin{rotate}{90}%
\footnotesize $-\frac{dW}{dx}/\left(\frac{e^4 T^2}{24 \pi}\right)$ 
\end{rotate}}}
\put(4,0.25){\makebox(0,0){\footnotesize $\hat{q}^{*}$}}
\end{picture}
\vspace{-1cm}
\caption{Energy loss as a function of $\hat{q}^{*}$ for $v=0.5$ and
$e=0.01$. The dashed line corresponds to the result from Braaten and Thoma.}
\label{fig:HPSv05}
\end{minipage} \hfill
\begin{minipage}[t]{.48\linewidth}
\includegraphics[width=0.8\linewidth]{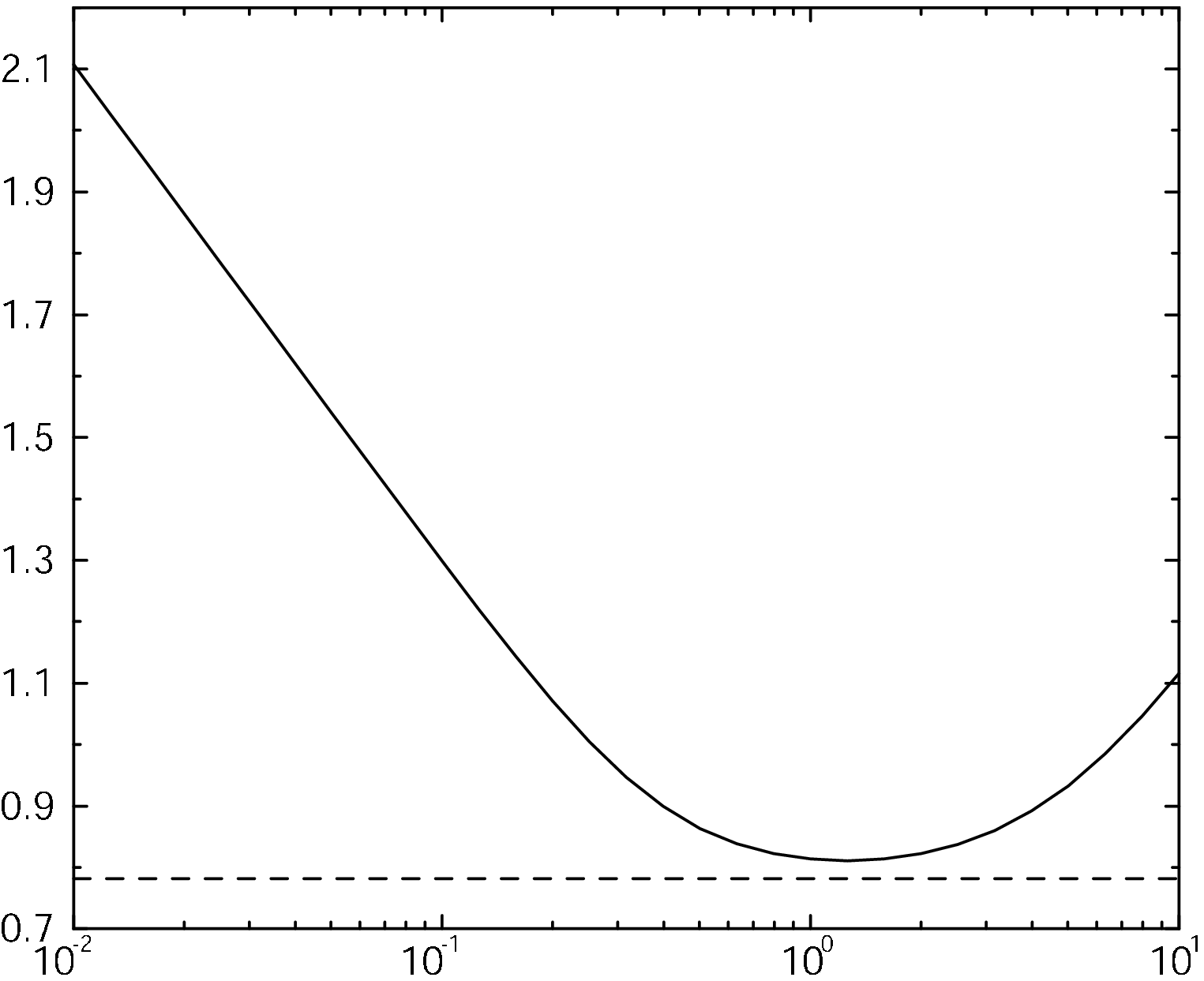}
\setlength{\unitlength}{1cm}
\begin{picture}(9,0)
\put(0.2,2.2){\makebox(0,0){\begin{rotate}{90}%
\footnotesize $-\frac{dW}{dx}/\left(\frac{e^4 T^2}{24 \pi}\right)$ 
\end{rotate}}}
\put(4,0.25){\makebox(0,0){\footnotesize $\hat{q}^{*}$}}
\end{picture}
\vspace{-1cm}
\caption{Energy loss as a function of $\hat{q}^{*}$ for $v=0.5$ and
$e=0.5$. The dashed line corresponds to the result from Braaten and Thoma. }
\label{fig:HPSv05_2}
\end{minipage}
\end{figure}

In principle $q^{*}$ should be an arbitrary quantity in the range $e\ll q^{*}/T 
\ll 1$ which is possible in the weak-coupling limit. Choosing $e=0.01$ and 
plotting $-dW/dx$ as a function of $\hat{q}^{*}=q^*/T$ at fixed $v$ and $T$ we find 
the result shown in Figure \ref{fig:HPSv05}. From this Figure we can see that 
the energy loss develops a plateau where the value is essentially identical to 
the result obtained by Braaten and Thoma; however, outside the plateau the function 
rises logarithmically with $\hat{q}^{*}$. For higher coupling ($e \agt 0.5$) the plateau 
shrinks rapidly to a minimum which can be seen in Figure \ref{fig:HPSv05_2}. 
We fix $\hat{q}^*$ by minimizing the energy loss with respect to $\hat{q}^*$. 

\begin{figure}
\includegraphics[width=9cm]{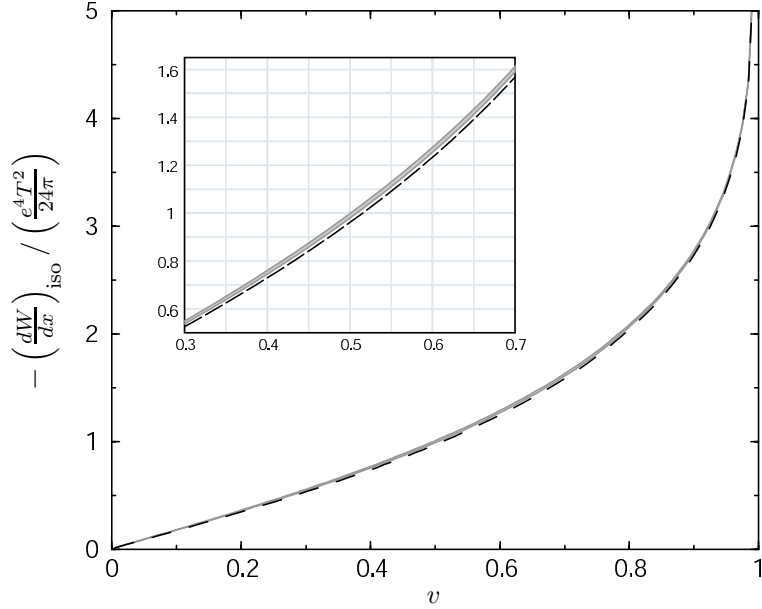}
\setlength{\unitlength}{1cm}
\begin{picture}(9,0)
\put(-0.5,3){\makebox(0,0){\begin{rotate}{90}%
$-\left(\frac{dW}{dx}\right)_{\rm iso}/\left(\frac{e^4 T^2}{24 \pi}\right)$ 
\end{rotate}}}
\put(4.7,0.25){\makebox(0,0){\footnotesize $v$}}
\end{picture}
\vspace{-0.3cm}
\caption{
Isotropic energy loss as a function of the electron velocity $v$ for $e=0.3$. 
Dashed line is the Braaten-Thoma result.  The grey band indicates the variation 
of the result obtained by varying $q^*$ by a factor of two around the central 
value.  Inset is an enlarged version of the region $0.3 < v < 0.7$.
}
\label{fig:isocompb}
\end{figure}

In Figure \ref{fig:isocompb} we compare our result with the result obtained by 
Braaten and Thoma for $e=0.3$.  The Braaten-Thoma result is shown as a dashed 
line and our result is shown as a grey band.  The band corresponds to the 
variation we obtain in our final result when varying the scale $\hat{q}^*$ by a 
factor of two around the central value (minimum).  Note that the choice of 
varying by a factor of two is completely arbitrary.  We could have easily chosen 
a smaller variation, however, we have chosen to be conservative here and vary 
$\hat{q}^*$ by a factor of two. As can be seen from this Figure, for $e=0.3$, the 
corrections to the Braaten-Thoma result are small; however, for larger couplings 
the band obtained by varying $\hat{q}^*$ by a factor of two can be quite large. 
The size of these bands gives us an estimate of the theoretical uncertainty 
present in the calculation.

\subsection{Anisotropic Case}

In the general anisotropic case we have to numerically evaluate the integrals 
contained in Eqs.~(\ref{Elosssoftfinal}) and (\ref{myelosshard3}) and add 
these to obtain the total energy loss
\beq
\left({\rm d} W \over {\rm d} x\right) = 
\left({\rm d} W \over {\rm d} x\right)_{\rm soft} + 
\left({\rm d} W \over {\rm d} x\right)_{\rm hard} \; .
\eeq
In addition to $v$ and $e$ the anisotropic energy loss also depends on
the anisotropy parameter $\xi$ and the angle of propagation with respect
to the anisotropy vector $\theta_n$.
Note that for every value of $v$, $e$, $\xi$, and $\theta_n$ there is a non-trivial 
consistency check between the hard and soft contributions, namely that 
the coefficients of $\log\hat{q}^*$ have to be equal and opposite for small-$\hat{q}^*$ 
and large-$\hat{q}^*$, respectively.  In every case that we have 
examined we find that these coefficients match to a part in $10^{-4}$ which 
can be further improved by increasing the target integration accuracy.

As in the isotropic case we find that for small coupling there is a plateau for 
$e \ll \hat{q}^* \ll 1$ and our final result is obtained by finding the minimum 
of the energy loss as a function of $\hat{q}^*$.  This corresponds to the 
point at which the result is minimally sensitive to $\hat{q}^*$.  As before, we 
can then obtain an estimate of the uncertainty in our prediction by varying 
$\hat{q}^*$ by a factor of two around the point of minimal sensitivity.  Note 
that using this method the central value of $\hat{q}^*$ always serves as a 
lower-bound on the prediction.

\begin{figure}
\begin{minipage}[t]{.48\linewidth}
\includegraphics[width=0.8\linewidth]{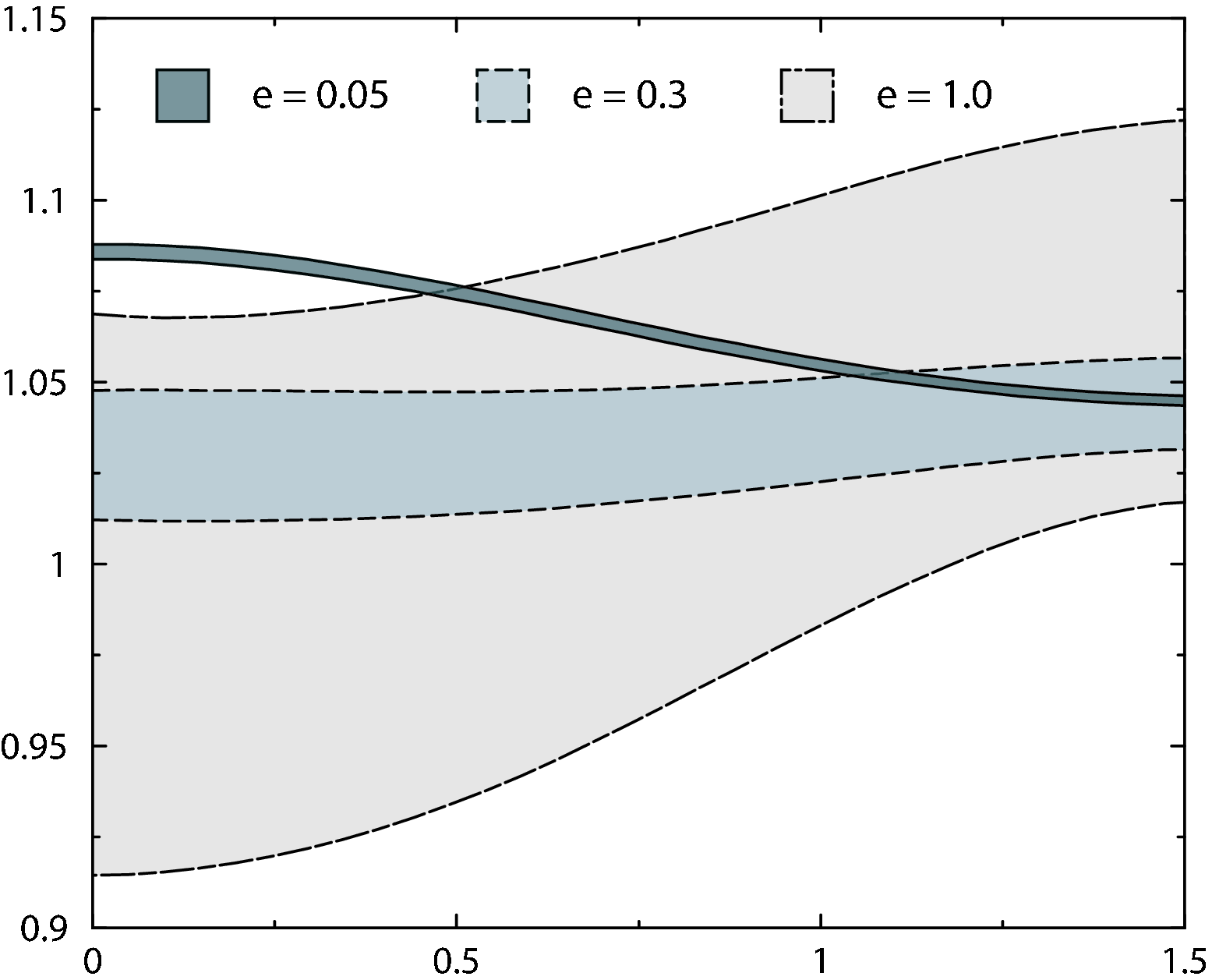}
\setlength{\unitlength}{1cm}
\begin{picture}(9,0)
\put(0.5,2.4){\makebox(0,0){\begin{rotate}{90}%
\footnotesize $\frac{dW}{dx}/\left(\frac{dW}{dx}\right)_{\rm iso}$
\end{rotate}}}
\put(4.2,0.25){\makebox(0,0){\footnotesize $\theta_n$}}
\end{picture}
\vspace{-1cm}
\caption{
Scaled energy loss as a function of the angle of propagation with respect to $\hat{\bf n}$
for $v=0.3$ and $\xi=1$.  
}
\label{fig:v0.3x1b}
\end{minipage} \hfill
\begin{minipage}[t]{.48\linewidth}
\includegraphics[width=0.8\linewidth]{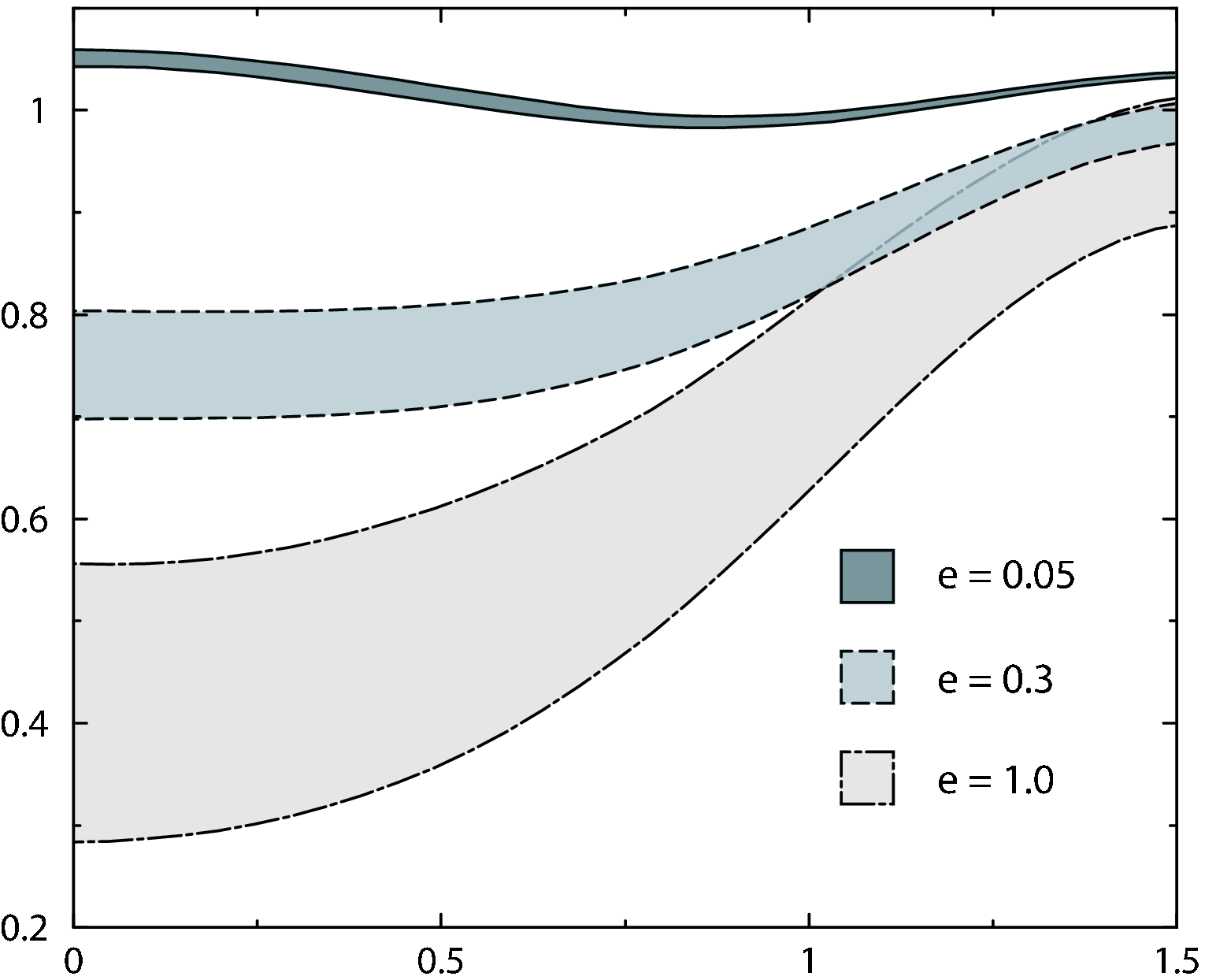}
\setlength{\unitlength}{1cm}
\begin{picture}(9,0)
\put(0.5,2.4){\makebox(0,0){\begin{rotate}{90}%
\footnotesize $\frac{dW}{dx}/\left(\frac{dW}{dx}\right)_{\rm iso}$
\end{rotate}}}
\put(4.2,0.25){\makebox(0,0){\footnotesize $\theta_n$}}
\end{picture}
\vspace{-1cm}
\caption{
Scaled energy loss as a function of the angle of propagation with respect to $\hat{\bf n}$
for $v=0.3$ and $\xi=10$.  
}
\label{fig:v0.3x10b}
\end{minipage}
\end{figure}

\begin{figure}
\begin{minipage}[t]{.48\linewidth}
\includegraphics[width=0.8\linewidth]{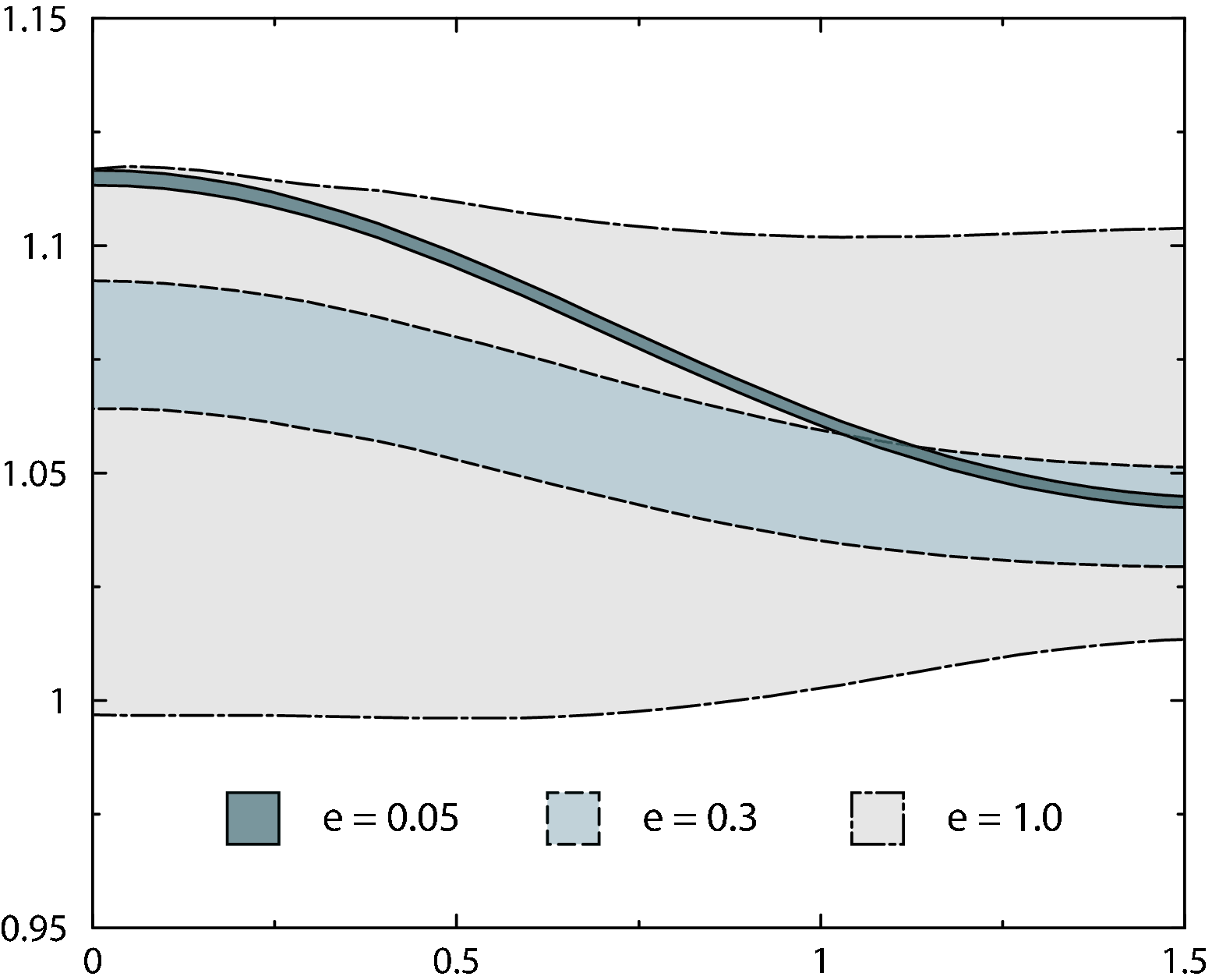}
\setlength{\unitlength}{1cm}
\begin{picture}(9,0)
\put(0.5,2.4){\makebox(0,0){\begin{rotate}{90}%
\footnotesize $\frac{dW}{dx}/\left(\frac{dW}{dx}\right)_{\rm iso}$
\end{rotate}}}
\put(4.2,0.25){\makebox(0,0){\footnotesize $\theta_n$}}
\end{picture}
\vspace{-1cm}
\caption{
Scaled energy loss as a function of the angle of propagation with respect to $\hat{\bf n}$
for $v=0.5$ and $\xi=1$.  
}
\label{fig:v0.5x1b}
\end{minipage} \hfill
\begin{minipage}[t]{.48\linewidth}
\includegraphics[width=0.8\linewidth]{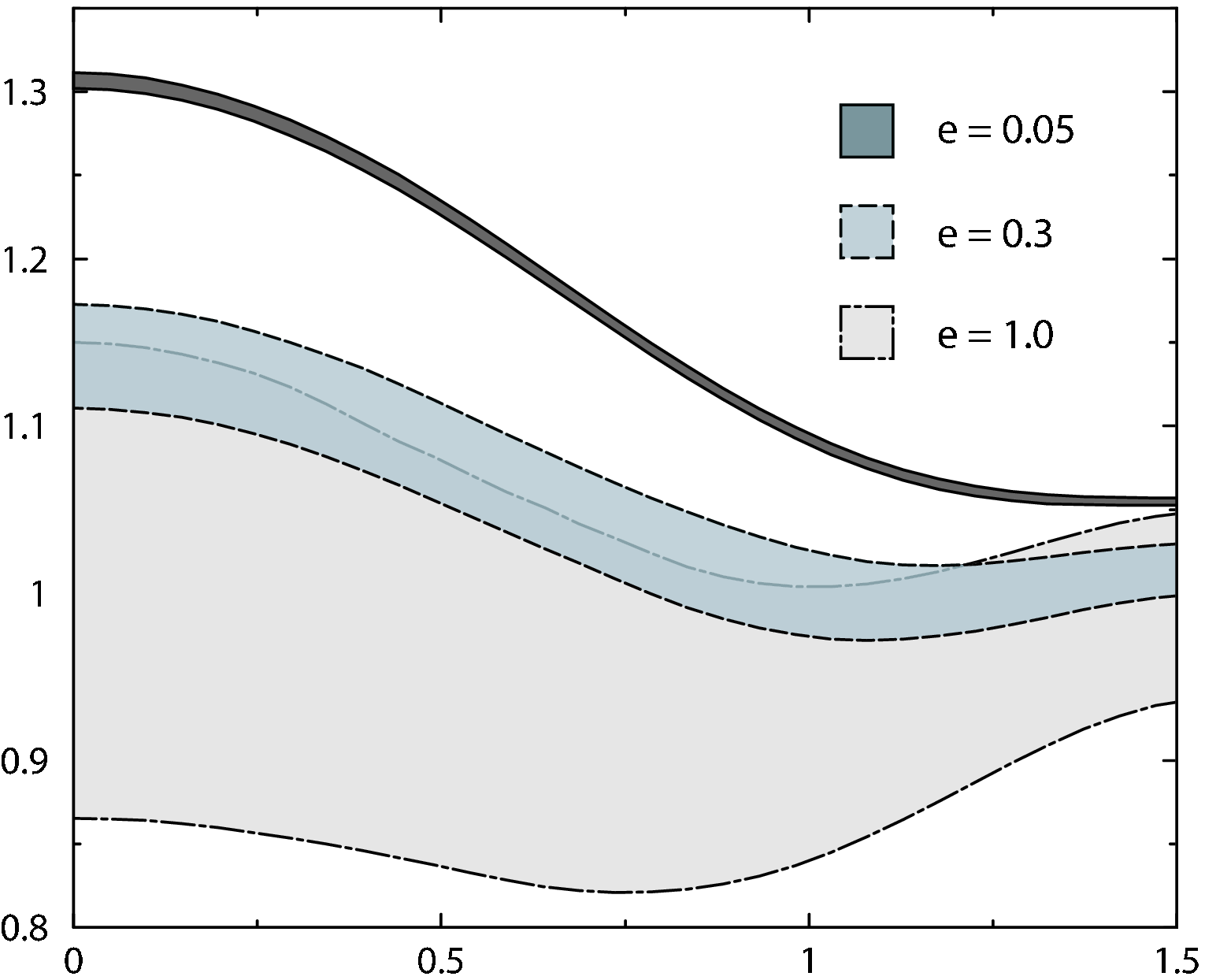}
\setlength{\unitlength}{1cm}
\begin{picture}(9,0)
\put(0.5,2.4){\makebox(0,0){\begin{rotate}{90}%
\footnotesize $\frac{dW}{dx}/\left(\frac{dW}{dx}\right)_{\rm iso}$
\end{rotate}}}
\put(4.2,0.25){\makebox(0,0){\footnotesize $\theta_n$}}
\end{picture}
\vspace{-1cm}
\caption{
Scaled energy loss as a function of the angle of propagation with respect to $\hat{\bf n}$
for $v=0.5$ and $\xi=10$.  
}
\label{fig:v0.5x10b}
\end{minipage}
\end{figure}

\begin{figure}
\begin{minipage}[t]{.48\linewidth}
\includegraphics[width=0.8\linewidth]{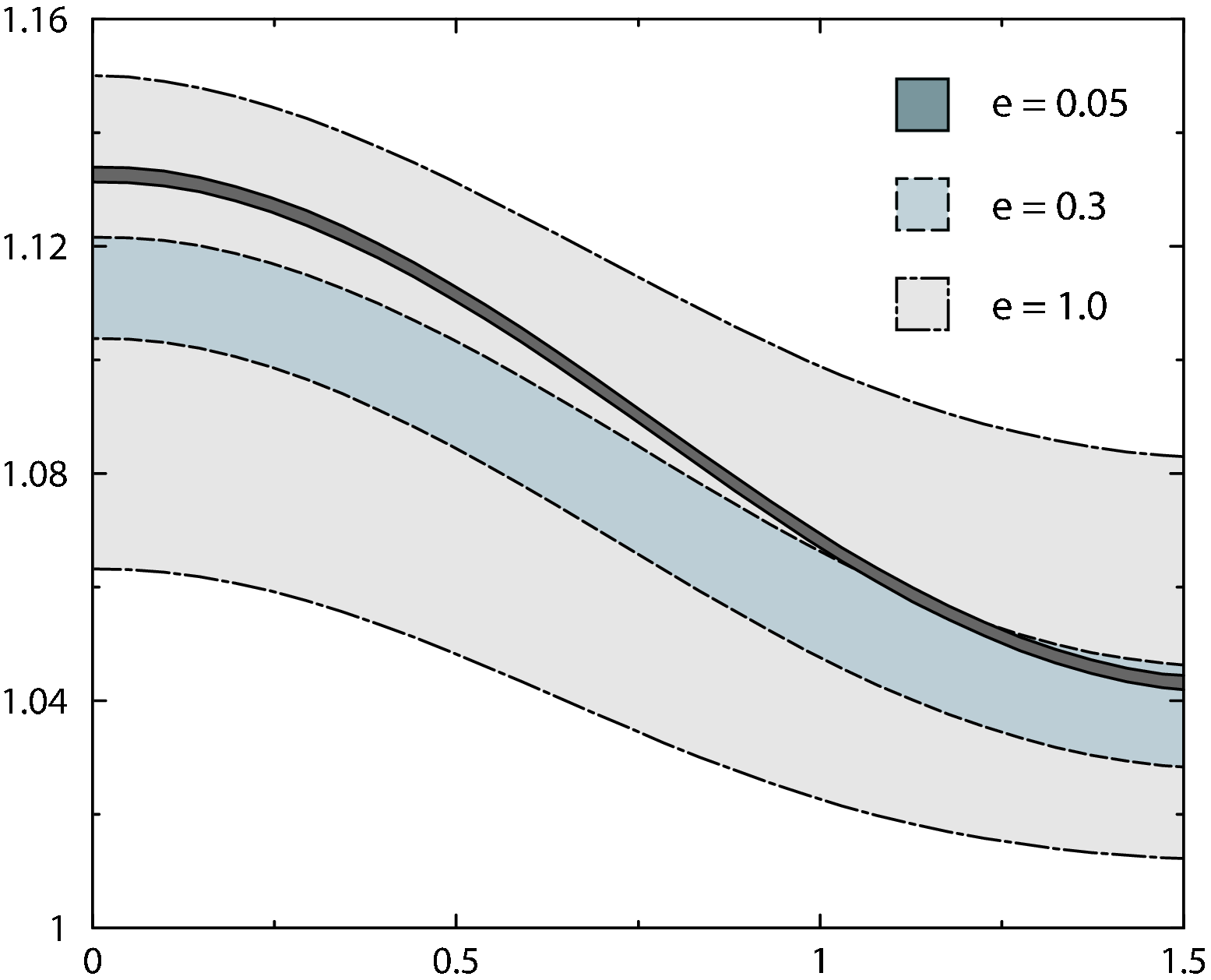}
\setlength{\unitlength}{1cm}
\begin{picture}(9,0)
\put(0.5,2.4){\makebox(0,0){\begin{rotate}{90}%
\footnotesize $\frac{dW}{dx}/\left(\frac{dW}{dx}\right)_{\rm iso}$
\end{rotate}}}
\put(4.2,0.25){\makebox(0,0){\footnotesize $\theta_n$}}
\end{picture}
\vspace{-1cm}
\caption{
Scaled energy loss as a function of the angle of propagation with respect to $\hat{\bf n}$
for $v=0.7$ and $\xi=1$.  
}
\label{fig:v0.7x1b}
\end{minipage} \hfill
\begin{minipage}[t]{.48\linewidth}
\includegraphics[width=0.8\linewidth]{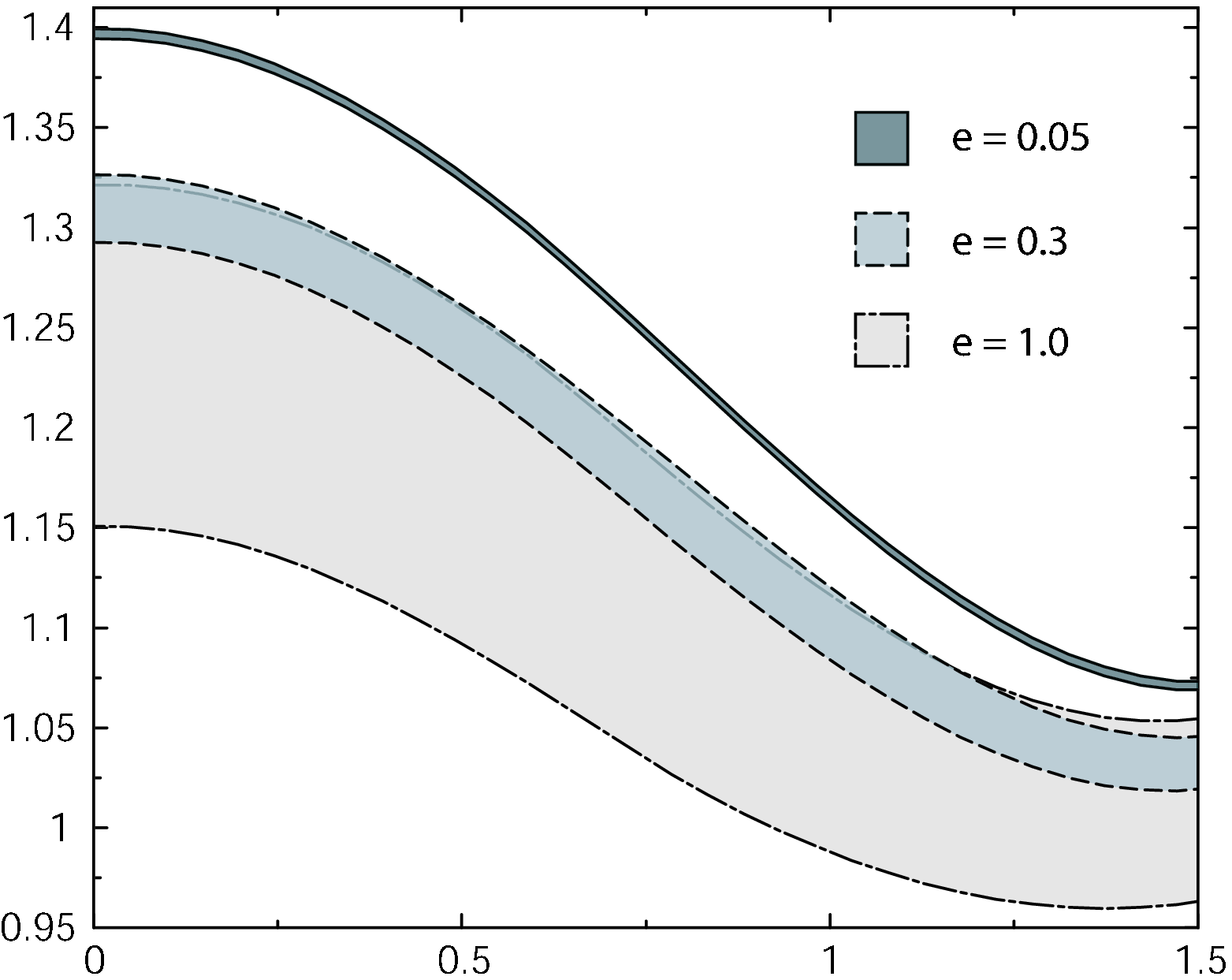}
\setlength{\unitlength}{1cm}
\begin{picture}(9,0)
\put(0.5,2.4){\makebox(0,0){\begin{rotate}{90}%
\footnotesize $\frac{dW}{dx}/\left(\frac{dW}{dx}\right)_{\rm iso}$
\end{rotate}}}
\put(4.2,0.25){\makebox(0,0){\footnotesize $\theta_n$}}
\end{picture}
\vspace{-1cm}
\caption{
Scaled energy loss as a function of the angle of propagation with respect to $\hat{\bf n}$
for $v=0.7$ and $\xi=10$.  
}
\label{fig:v0.7x10b}
\end{minipage}
\end{figure}

In Figures \ref{fig:v0.3x1b}-\ref{fig:v0.7x10b} we plot the $\theta_n$-dependence 
of the resulting anisotropic energy loss for $v=\{0.3,0.5,0.7\}$, 
$e=\{0.05,0.3,1.0\}$, and $\xi=\{1,10\}$ scaled by the corresponding isotropic 
energy loss.  For $e=0.05$ we see that for all three values of the velocity 
the energy loss is larger along the direction of the anisotropy then 
transverse to it.  For the physical value of electromagnetic coupling constant, 
$e=0.3$, we find that directional dependence of the energy loss depends on the 
velocity and the strength of the anisotropy. For $v=0.3$ and $\xi=1$ the result 
is almost independent of the direction of propagation but if the anisotropy is 
stronger ($\xi=10$) then the energy loss is less in the direction of the 
anisotropy than transverse to it.  Note that this is consistent with the results 
obtained in the small-$\xi$ limit in Sections \ref{softsmallxibehavior} and 
\ref{hardsmallxibehavior}.  For larger velocities we find that the energy loss 
along the direction of the anisotropy is higher than transverse to it; however, 
if the coupling constant is increased to a large enough value then the trend 
will reverse as was the case for smaller velocities.  Note that for $v \agt 0.5$ 
the coupling constant required to reverse the trend is extremely large.  

In the most extreme case shown here in Figure \ref{fig:v0.7x10b} for which 
$v=0.7$ and $\xi=10$ we see that for all values of the coupling constant shown 
the energy loss varies by 20-25\% depending on whether the fermion is propagating 
along the direction of the anisotropy or transverse to it.  For larger values of 
the velocity and anisotropy parameter the directional dependence of the energy 
loss increases.  

\begin{figure}
\begin{minipage}[t]{.48\linewidth}
\includegraphics[width=0.8\linewidth]{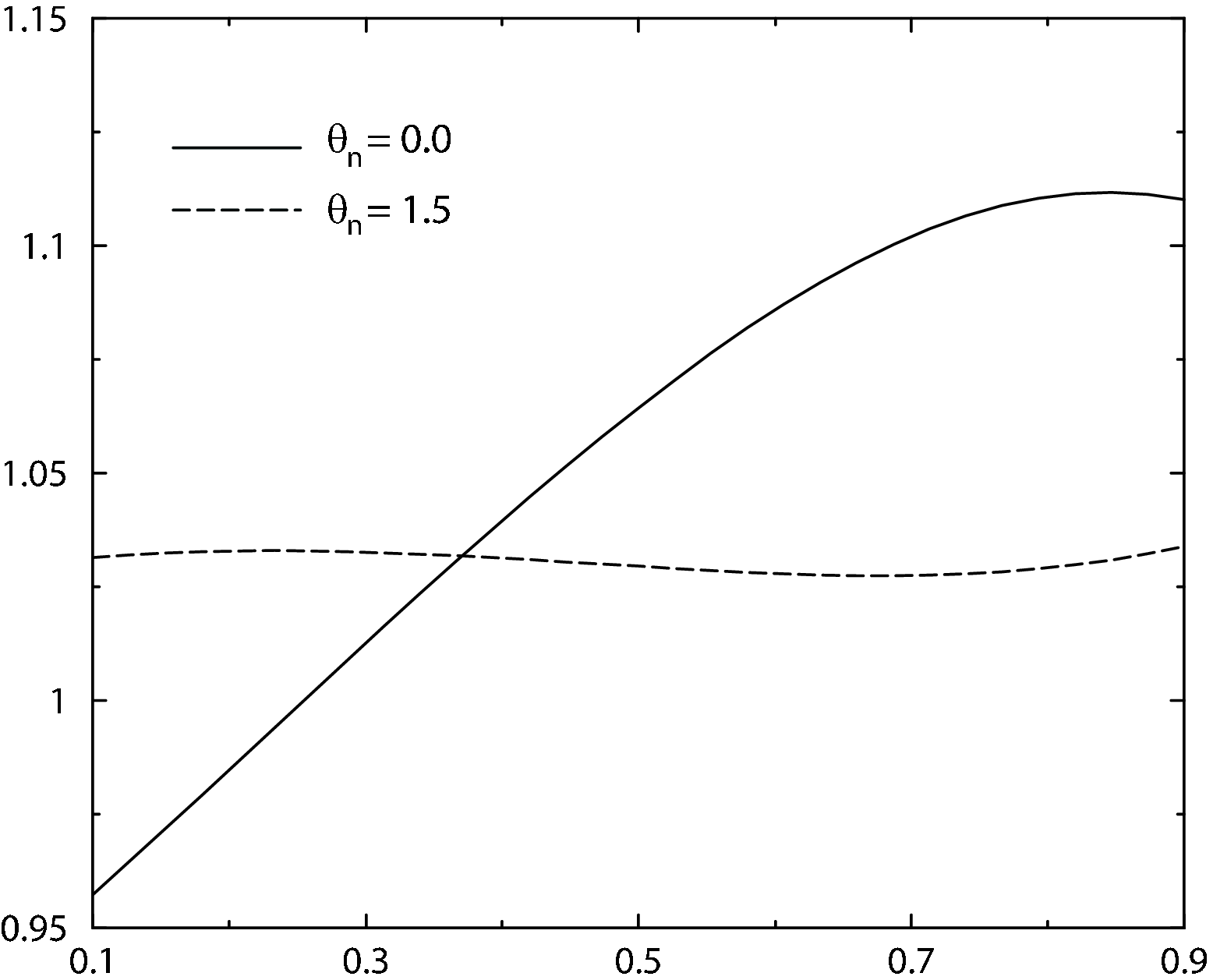}
\setlength{\unitlength}{1cm}
\begin{picture}(9,0)
\put(0.5,2.4){\makebox(0,0){\begin{rotate}{90}%
\footnotesize $\frac{dW}{dx}/\left(\frac{dW}{dx}\right)_{\rm iso}$
\end{rotate}}}
\put(4.2,0.25){\makebox(0,0){\footnotesize $v$}}
\end{picture}
\vspace{-1cm}
\caption{
Energy loss for angles $\theta_n=\{0,1.5\}$ as a function of $v$ for
$\xi=1$ and $e=0.3$.
}
\label{fig:vdep}
\end{minipage} \hfill
\begin{minipage}[t]{.48\linewidth}
\includegraphics[width=0.8\linewidth]{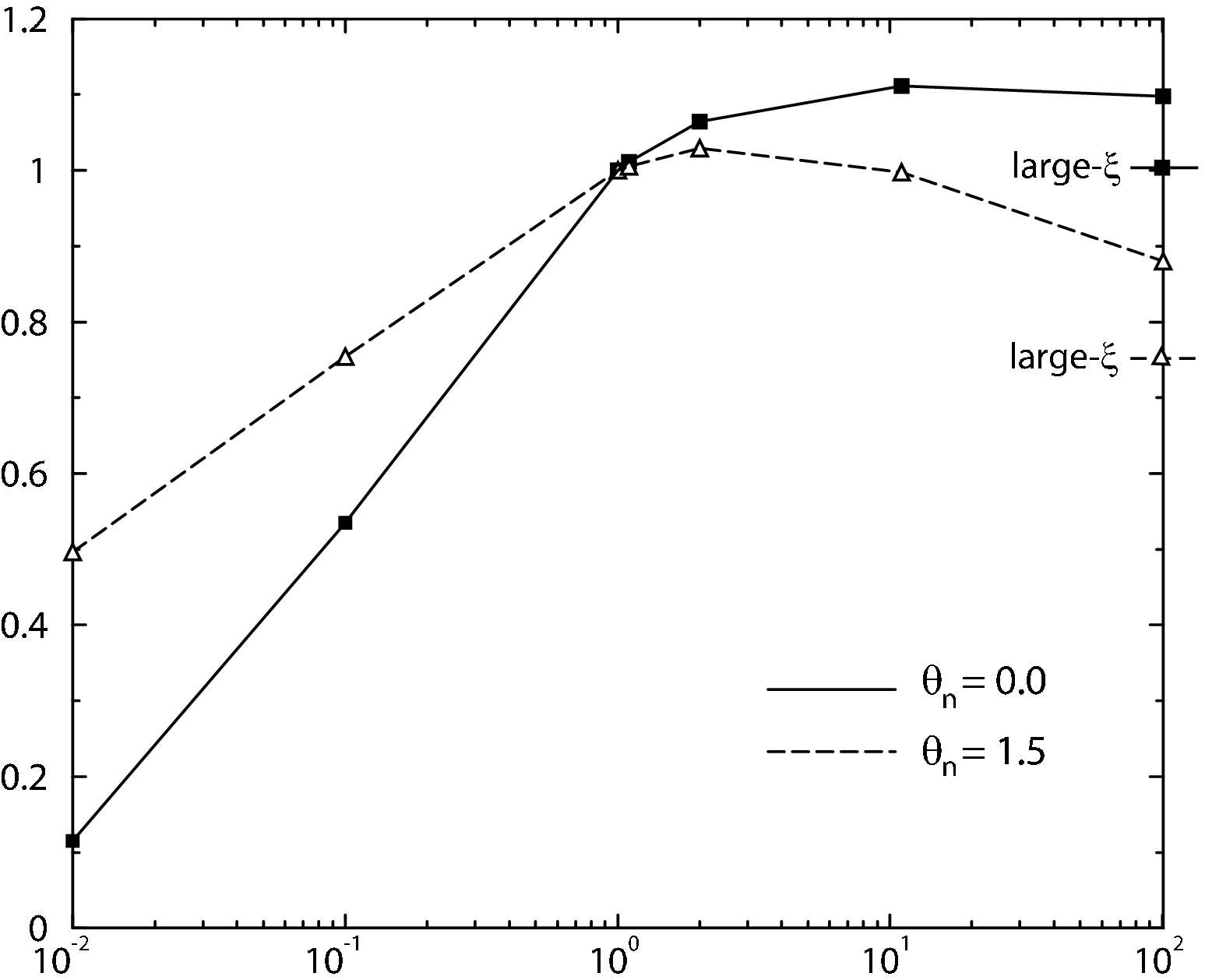}
\setlength{\unitlength}{1cm}
\begin{picture}(9,0)
\put(0.5,2.4){\makebox(0,0){\begin{rotate}{90}%
\footnotesize $\frac{dW}{dx}/\left(\frac{dW}{dx}\right)_{\rm iso}$
\end{rotate}}}
\put(4,0.3){\makebox(0,0){\footnotesize $1+\xi$}}
\end{picture}
\vspace{-1cm}
\caption{
Energy loss for angles $\theta_n=\{0,1.5\}$ as a function of $\xi$ for
$v=0.5$ and $e=0.3$.  Lines on right hand side of the plot frame indicate
results obtained in the large-$\xi$ limit.
}
\label{fig:xidep}
\end{minipage} 
\end{figure}

In Figure \ref{fig:vdep} we plot the dependence of the energy loss normalized to 
the isotropic result for $\xi=1$, $e=0.3$, and $\theta_n=\{0,1.5\}$ as a function 
of the velocity $v$.  From this Figure we see that for fixed $e$ and $\xi$ the 
angular dependence (forward-peaked versus transversely-peaked) changes as the 
velocity is increased with the energy loss being transversely-peaked for small 
velocities and forward-peaked for large velocities.  For the largest velocity 
shown in this Figure we see that for $\xi=1$ the effect is on the order of 10 
\%.  

In Figure \ref{fig:xidep} we plot the dependence of the energy loss for 
$v=0.5$, $e=0.3$, and $\theta_n=\{0,1.5\}$ as a function of $\xi$.  
Additionally, we have included results obtained by taking the large-$\xi$ limit
by lines on the right hand side of the plot frame.  As can be seen from this Figure 
for $\xi>0$ we find that the difference between the forward and transverse 
energy loss saturates at approximately 25\% in the large $\xi$ limit.  For
$\xi<0$ we see that for both angles shown the energy loss approaches zero but
the difference between the forward and transverse energy loss is still 
quite large at $\xi=-0.99$.  Therefore, it is possible that the presence of 
momentum-space anisotropies in the electron distribution function could lead
to a rather significant experimental effect. 

\section{Conclusions}
\label{conclusions}

In this paper we have derived integral expressions for the collisional energy 
loss of a heavy fermion propagating through a QED plasma for which the electron 
distribution function is anisotropic in momentum space.  We then numerically 
evaluated the resulting integrals and studied the dependence of the heavy 
fermion energy loss on the angle of propagation, degree of momentum-space 
anisotropy, and coupling constant.  We have shown that the techniques used in 
the isotropic case can be straightforwardly extended to the anisotropic case. 
We have also discussed how problems due to unstable soft photonic modes
are avoided in the case of the energy loss.

As a side result we demonstrated that in the isotropic case the canonical 
result from Braaten and Thoma \cite{BT:1991} has a correction which comes from 
the dependence of the result on the scale introduced to separate the soft and 
hard contributions to the energy loss.  The residual dependence on the 
separation scale was then used to estimate the theoretical uncertainty of the 
resulting calculation for both the isotropic and anisotropic case.  

Our final results indicate that for anisotropic QED plasmas there can be a 
significant directional dependence of the energy loss for highly-relativistic 
fermions if there is a strong momentum-space anisotropy present in the electron 
distribution function.  It should be mentioned that we have assumed here that
the heavy fermion is infinitely massive.  In the realistic case when the fermion
mass is on the order of the muon mass the result obtained here should be reliable
for velocities $0.6 \alt v \alt 0.95$ \cite{BT:1991}.

We have concentrated here on the collisional energy loss in QED.  In order to
extend the result to QCD there will be a modification of the hard energy loss
due to the fact that the fermion-boson scattering diagrams do not cancel against
one another in this case.  Besides this, the only modification required will be
to adjust the Debye mass so that it corresponds to the QCD Debye mass.  One
complication will come from the fact that the quark and gluon distribution
functions can, in general, have different momentum-space anisotropies; however,
there should be an independent matching of logarithms coming from the quark 
and gluonic sectors.  This work is currently in progress.

\section*{Acknowledgments}
M.S. and P.R. would like to thank A.~Ipp, A.~Rebhan, and M.~Waibel for discussions.  
M.S. was supported by the Austrian Science Fund Project No. M689.
P.R. was supported by the Austrian Science Fund Project No. P14632.

\bibliography{bsample}
\bibliographystyle{utphys}

\end{document}